\documentclass[a4paper,fleqn,usenatbib]{mnras}

\usepackage{newtxtext,newtxmath}

\usepackage[T1]{fontenc}
\usepackage{ae,aecompl}

\usepackage{graphicx}	
\usepackage{amsmath}	
\usepackage{amssymb}	
\usepackage{siunitx}
\usepackage{booktabs}
\usepackage{xspace}

\newcommand{\kepler}{{\textsc{Kepler}}\xspace}

\title[Matching X-ray Burst Observations with Models]{A Bayesian Approach to Matching Thermonuclear X-ray Burst Observations with Models}

\author[Goodwin et al.]{
A. J. Goodwin$^{1,2}$,\thanks{E-mail: ajgoodwin.astro@gmail.com}
\ D. K. Galloway$^{1,2}$,
\ A. Heger$^{1,2,3}$,
\ A. Cumming$^{4}$,
\ and Z. Johnston$^{1,2}$
\\
$^{1}$School of Physics \& Astronomy, Monash University, Clayton VIC
3800, Australia\\
$^{2}$also Monash Centre for Astrophysics (MoCA)\\
$^{3}$Tsung-Dao Lee Institute, Shanghai 200240, China\\
$^{4}$Department of Physics and McGill Space Institute, McGill University, 3550 rue University, Montreal, QC, H3A 2T8, Canada\\
}

\date{Accepted 2019 September 16. Received 2019 September 05; in original form 2019 July 01}

\pubyear{2019}

\begin{document}
\label{firstpage}
\pagerange{\pageref{firstpage}--\pageref{lastpage}}
\maketitle

\begin{abstract}

We present a new method of matching observations of Type I (thermonuclear) X-ray bursts with models, comparing the predictions of a semi-analytic ignition model with X-ray observations of the accretion-powered millisecond pulsar \texttt{SAX J1808.4--3658} in outburst. We used a Bayesian analysis approach to marginalise over the parameters of interest and determine parameters such as fuel composition, distance/anisotropy factors, neutron star mass and neutron star radius. Our study includes a treatment of the system inclination effects, inferring that the rotation axis of the system is inclined $\left(69^{+4}_{-2}\right)^\circ$ from the observers line of sight, assuming a flat disc model. This method can be applied to any accreting source that exhibits Type I X-ray bursts.  We find a hydrogen mass fraction of $0.57^{+0.13}_{-0.14}$ and CNO metallicity of $0.013^{+0.006}_{-0.004}$ for the accreted fuel is required by the model to match the observed burst energies, for a distance to the source of $3.3^{+0.3}_{-0.2}\,\mathrm{kpc}$.  We infer a neutron star mass of $1.5^{+0.6}_{-0.3}\,\mathrm{M}_{\odot}$ and radius of $11.8^{+1.3}_{-0.9}\,\mathrm{km}$ for a surface gravity of $1.9^{+0.7}_{-0.4}\times10^{14}\,\mathrm{cm}\,\mathrm{s}^{-2}$ for \texttt{SAX J1808.4--3658}. 

\end{abstract}

\begin{keywords}
X-rays: bursts -- X-rays: binaries -- pulsars: individual: \texttt{SAX J1808.4--3658}
\end{keywords}

\section{Introduction}
Accretion-powered millisecond pulsars (AMSPs) are neutron stars with weak ($\sim10^8$ G) magnetic fields that are accreting matter from a low mass companion star \citep[e.g.,][]{lewin1993,Chakrabarty1998,Wijnands1998,Patruno2012,Galloway2017review}. These systems are typically transient X-ray sources, and exhibit coherent X-ray pulsations as well as Type I thermonuclear X-ray bursts periodically \citep[e.g.,][]{Wijnands2004}. A source goes into outburst when the accretion disc around the neutron star, formed by accretion of matter onto the companion star, reaches a critical state that causes matter to transfer directly onto the neutron star and ignites thermonuclear bursts \citep[e.g.,][]{Strohmayer2006,Galloway2017review}.
Type I X-ray bursts are the most frequently observed thermonuclear powered outbursts in nature and provide a unique insight into the nuclear reactions that can occur in extreme environments. They are characterised by a sudden increase in the X-ray flux of a source in outburst \citep[e.g.,][]{galloway2008}. They can be observed from neutron stars in binary systems, and are thought to be caused by unstable ignition of hydrogen or helium on the surface of an accreting neutron star \citep[e.g.,][]{Strohmayer2006}.

The challenge inherent in understanding low-mass neutron star binary systems is estimating system parameters, including the composition of the fuel that drives the thermonuclear burst; and the distance/inclination of the source, which cannot be directly observed as they are kilometre-sized objects at kilo-parsec distances. These parameters are an essential ingredient in modelling and understanding the dense matter of neutron stars and the physical processes behind thermonuclear bursts.  Applications range from stellar evolution (with fuel composition of the bursts enabling constraints on the evolution of the companion star and population synthesis) to constraining proton- and carbon-rich nuclear reactions that are difficult to replicate in Earth-bound laboratories. 

Neutron stars are the densest directly observable stellar objects known. The study of neutron stars through various electromagnetic bands has seen much progress in recent years, furthering the understanding of high-density cold matter and the dense equation of state \citep[e.g.][]{Miller2019,Degenaar2018,Ozel2016} The recent observation of gravitational waves from merging neutron stars \citep{Abbott2017} has also seen an increase in efforts to constrain neutron star properties, such as the neutron star mass, radius, spin, and equation of state \citep[e.g.,][]{Abbott2018,Annala2018,Radice2018}. Constraining these properties is fundamental in the understanding of ultra-dense matter, such as the material that makes up a neutron star, as well as providing additional constraints in the search for more gravitational waves. \\
\\

Observations of AMSPs that exhibit Type I X-ray bursts are becoming increasingly more accessible, with the Multi-INstrument Burst Archive (MINBAR)\footnote{https://burst.sci.monash.edu/minbar/} providing an extensive collection of processed and anlaysed X-ray burst data. Observations are limited to the X-ray activity of these sources, providing measurements of the accretion flux, burst flux, and recurrence time of bursts.

Efforts to model burst sources have evolved from early models \citep[e.g.,][]{Fujimoto1987,taam1980,Paradijs1988} that focused on the correlation between burst energies and recurrence times, to varying success, to more recent models \citep[e.g.,][]{cumming2003,galloway2004,woosley2004} which have a heavier focus on the nuclear physics driving the bursts, predicting fuel compositions and accretion rates by implementing more detailed models of the nuclear reaction rates that produce the observed energy generation rates. 

These observations and models motivate an effort to combine the data obtained from both, to obtain a deeper understanding of the thermonuclear processes that occur during an X-ray burst, as well as identifying where current models are failing to predict observed properties \citep[e.g.,][]{Galloway2017}. The most well-studied transients, such as the accretion-powered millisecond pulsar \texttt{SAX J1808.4--3658}, provide crucial test cases for numerical models of thermonuclear bursts and can be used as standard cases in modelling studies. 

\citet{galloway2006} showed that the bursts from \texttt{SAX J1808.4--3658} were consistent with ignition in an almost pure He environment. The long recurrence times of $\approx $1 day allow enough time for hydrogen to be depleted by thermally-stable hot CNO burning between bursts. By modelling the burst properties, they were able to constrain the hydrogen fraction in the accreted layer, and showed that the observed recurrence times imply a relation between the hydrogen fraction and fraction of CNO elements. As such, \texttt{SAX J1808.4--3658} provides the best example of pure helium ignited bursts, and validates the theory in which hydrogen depletes before ignition of the X-ray burst in a helium environment.

In this study we develop a new method for matching burst observations with a semi-analytic ignition model, extending the previous work of \citet{galloway2006}, to determine system parameters such as fuel composition, neutron star mass and radius, distance, and inclination. We use a Markov Chain Monte Carlo approach to find the best-fitting system parameters, with uncertainties. In Section \ref{sec:methods} we describe the semi-analytic model used in this study, compare it to detailed time-dependent simulations, and describe the MCMC implementation. In Section \ref{sec:results} we report the parameter limits predicted by the model, and describe how well our model matches the observations. Finally, in Section \ref{sec:conclusion}, we summarise the results and give our conclusions.

\section{Methods}\label{sec:methods}

We used the ignition model of \citet{Cumming2000} (hereafter called {\sc settle}) to generate burst sequences, which were then compared to \textit{Rossi X-ray Timing Explorer} (RXTE) Proportional Counter Array (PCA) telescope observations of the 2002 outburst of \texttt{SAX J1808.4--3658} using a Markov Chain Monte Carlo (MCMC) algorithm. We obtained the `best' model run that matched the observations, providing predictions for parameters such as fuel composition, distance, anisotropies, neutron star mass and radius, that cannot be measured otherwise. We used the python implementation of MCMC, {\sc emcee} \citep[][]{Foreman2013}. RXTE observations were obtained from archival data available in MINBAR and shown in Table \ref{tab:1808_observations}, updated since the initial analysis of this outburst by \citet{galloway2006}. Observed properties of the bursts include times, persistent flux, burst fluence and $\alpha$ values. Here, $\alpha$ is defined as the ratio of integrated persistent flux to the fluence, given by

\begin{equation}
\label{eq:alpha}
        \alpha = \frac{F_{\mathrm{p}} c_{\mathrm{bol}} \Delta t}{E_{\mathrm{b}}} 
\end{equation}

where $E_{\mathrm{b}}$ is the burst fluence, $\Delta t$ is the recurrence time, $F_{\mathrm{p}}$ is the persistent flux, and $c_{\mathrm{bol}}$ is the bolometric correction. For $c_{\mathrm{bol}}$ we adopted a constant value of $2.12$, as in \citet{galloway2006}.

\begin{table}
    \begin{centering}
	\caption{Observations of the 2002 October outburst of \texttt{SAX J1808.4--3658}}
	\setlength{\tabcolsep}{2pt}
	\label{tab:1808_observations}
	\begin{tabular}{lcccr}
    \hline
        No. & Burst start time & Peak Flux & Fluence & $\alpha$ \\
        &(MJD)&(10$^{-9}\,\mathrm{erg}\,\mathrm{cm}^{-2}\,\mathrm{s}^{-1}$)&($10^{-6}\,\mathrm{erg}\,\mathrm{cm}^{-2}$)&\\
        \hline
        1 & 52562.4136 & 163$\pm$3 & 2.62$\pm$0.02& -- \\
        2 & 52564.3051 & 177$\pm$3 & 2.65$\pm$0.01& 107$\pm$2 \\
        3 & 52565.1843 & 179$\pm$3 & 2.99$\pm$0.01& 118$\pm$2 \\
        4 & 52566.4268 & 177$\pm$3 & 3.46$\pm$0.02 & 128$\pm$2 \\
        \hline
    \end{tabular}
    \end{centering}
    \small{\\\textbf{Notes:} The $\alpha$ values are calculated between bursts using Equation \ref{eq:alpha}, assuming 2 bursts occurring between Bursts 1 and 2. Fluxes reported are bolometric fluxes and uncertainties are 1$\sigma$ confidence levels.}
\end{table}

In order to determine the accretion rate at the time of a burst, we used persistent flux measurements of the outburst from the PCA and linearly interpolated the values between measurements either side of the burst. This enabled us to obtain a value at the time of each burst, from which an accretion rate was inferred using

\begin{equation}
\label{eq:fptomdot}
    \dot{m} = c_{\mathrm{bol}}F_{\mathrm{p}} r_{\mathrm{1}}
\end{equation}
where $F_{\mathrm{p}}$ is the persistent flux, $\dot{m}$ is the accretion rate, and $r_{\mathrm{1}}$ is the scaling factor between the accretion rate and the persistent flux, described in Section \ref{sec:mcmc}.

\subsection{{\sc settle}}

{\sc settle} is a semi-analytic ignition model developed by \citet{Cumming2000} that integrates the thermal profile of a settling atmosphere to find ignition conditions for a single burst, from which the burst train can be predicted. It applies a one-zone ignition criterion to simple models of the accumulating layer, which allows a survey of parameter space whilst also enabling estimation of the ignition depth. Researchers have applied {\sc settle} to observations of regular Type I bursters, such as \texttt{4U 1820--30} \citep{cumming2003}, \texttt{GS 1826--24} \citep{cumming2003}, and \texttt{SAX J1808.4--3658} \citep{galloway2006}. The {\sc settle} model assumes the accumulating fuel layer is heated by hot CNO hydrogen burning within the layer, and a constant flux from deeper layers. It does not include heating from energy released in a previous burst, or the possibility of ignition of unburnt fuel within the ashes of a previous burst (commonly referred to as thermal and compositional inertia; \citealt{woosley2004}; see Section~\ref{sec:comparison} for corrections we applied to {\sc settle} to account for some of these simple assumptions). Assuming spherical symmetry, {\sc settle} does not address complex physics that is currently poorly understood, such as how the burning front spreads across the star after ignition occurs, what determines the number of hotspots on the surface of the star, or the cause of asymmetry near the end of bursts as the material is cooling \citep[e.g.,][]{yuri2019}. 

{\sc settle} adopts a neutron star mass and radius (e.g., $M_{\mathrm{NS}} = 1.4\,\mathrm{M}_{\odot}$ and $R_{\mathrm{NS}} = 11.2\,\mathrm{km}$) which then sets the gravitational redshift and surface gravity, where the surface gravity, $g$, in the X-ray burst layer is given by

\begin{equation}
\label{eq:grav}
\centering
    g = \frac{G M_{\mathrm{NS}}}{R_{\mathrm{NS}}^2} \cdot (1+z)
\end{equation}

where G is the Gravitational constant, $z$ is the gravitational redshift, $M_{\mathrm{NS}}$ is the gravitational neutron star mass and $R_{\mathrm{NS}}$ is the neutron star radius in the observer frame.

An X-ray burst is triggered when helium burning becomes unstable at the base of the accumulated layer, at a temperature of $\approx 2 \times 10^8$ K and a density range of $\sim 10^5$--$10^6\,\mathrm{g}\,\mathrm{cm}^{-3}$. The temperature profile of the accumulating layer of hydrogen and helium is calculated given the accretion rate, and the accumulating layer is allowed to build up until the one-zone condition for a thermal runaway is satisfied at the base of the layer. This condition is met when $\mathrm{d}\epsilon_{3\alpha}/\mathrm{d}T > \mathrm{d}\epsilon_{\mathrm{cool}}/\mathrm{d}T$, where $\epsilon_{3\alpha}$ is the energy production rate of helium burning (via the triple alpha reaction) and $\epsilon_{\mathrm{cool}}$ is the effective local cooling rate. At this point, a thermal instability occurs and the conditions for thermonuclear runaway are met. The temperature profile at ignition is determined such that $\mathrm{d}\epsilon_{3\alpha}/\mathrm{d}T = \mathrm{d}\epsilon_{\mathrm{cool}}/\mathrm{d}T$ at the base of the ignition column. 

The composition of the ignition column at ignition depends on how much hydrogen has burned before the ignition conditions were met, during the accumulation of the fuel. The amount of hydrogen burned is determined by the local accretion rate, since hydrogen is assumed to burn steadily at a constant rate via the hot CNO cycle. The rate at which hydrogen burns is thus given by 

\begin{equation}
\label{eq:Hrate}
    \frac{\mathrm{d}Y_{\mathrm{^1H}}}{\mathrm{d}t} = \frac{4\ln{2}\,Y_{\mathrm{CNO}}}{t_{\mathrm{\frac{1}{2}, CNO}}}\;,
\end{equation}
where $Y_{\mathrm{CNO}}$ is the parts per nucleon of CNO elements and $t_{\mathrm{\frac{1}{2}, CNO}}$ is the cycle half life, i.e., the sum of the half lives of the isoptopes limiting the cycle.  In {\sc settle}, we assume the CNO elements are the $\beta$-limited CNO cycle isotopes, $^{14}$O and $^{15}\mathrm{O}$, in their cycle equilibrium abundance values and $Y_{\mathrm{CNO}}$ can be computed from 

\begin{equation}
    \label{eq:Ycno}
    Y_{\mathrm{CNO}} = \frac{Z_{\mathrm{CNO,\beta}}}{\bar{A}_{\mathrm{CNO,\beta}}}
\end{equation}
where $\bar{A}_{\mathrm{CNO,\beta}}$ is the average mass number of the CNO isotopes in the $\beta$-limited CNO cycle, given by 

\begin{equation}
    \label{eq:Abar}
\bar{A}_{\mathrm{CNO,\beta}} = \frac{t_{\mathrm{\frac{1}{2}, ^{15}O}} A_{\mathrm{^{15}O}} + t_{\mathrm{\frac{1}{2}, ^{14}O}} A_{\mathrm{^{14}O}}}{t_{\mathrm{\frac{1}{2}, ^{15}O}} + t_{\mathrm{\frac{1}{2}, ^{14}O}}}\;,
\end{equation}
where $A_i$ is the mass number of species $i$, $t_{\frac{1}{2}, ^{15}\mathrm{O}} = 122\,\mathrm{s}$ and $t_{\frac{1}{2}, ^{14}\mathrm{O}} = 71\,\mathrm{s}$.

The ratio of the parts per nucleon of $^{14}$O to $^{15}$O is given by the ratio of their half lives. Thus the mass fractions $Z_{\mathrm{14}} = 0.352\,Z_{\mathrm{CNO}}$ and $Z_{\mathrm{15}} = 0.648\,Z_{\mathrm{CNO}}$, giving $\bar{A}_{\mathrm{CNO},\beta}=14.653\,\mathrm{g}\,\mathrm{mol}^{-1}$. 

The energy production rate of hydrogen burning via the CNO cycle, $\epsilon_{\mathrm{H}}$, is given by
$$
\epsilon_{\mathrm{H}} = \frac{\mathrm{d}}{\mathrm{d}t} Y_{\mathrm{^1H}}\, E_{\mathrm{H}}\;,
$$ 
where $E_{\mathrm{H}}$ is the energy produced from burning $4\,^1\mathrm{H}$ to $^4\mathrm{He}$.  $E_{\mathrm{H}}$ is given by the change in mass excess less the energy carried away by neutrinos, $E_{\mathrm{H}}=6.03\times10^{18}\,\mathrm{erg}\,\mathrm{g}^{-1}$.  Thus, $\epsilon_{\mathrm{H}}$ is given by  
\begin{equation}
\label{eq:epsH}
    \epsilon_\mathrm{H} = 5.94 \times 10^{13}\, \frac{Z_{\mathrm{CNO,\beta}}}{0.01}\mathrm{\quad erg\,g^{-1}\,s^{-1}}\;,
\end{equation}
where $Z_{\mathrm{CNO,\beta}}$ is the mass fraction of CNO nuclei in the $\beta$-limited CNO cycle, computed as outlined above.

We also compare to the Sun as a reference point with its own, different, distribution of the CNO isotopes, with $\bar{A}_{\mathrm{CNO},\odot}=14.688$, such that a given CNO metallicity translates into a slightly different value of $Y_\mathrm{CNO}$.  Using the solar abundance ratios for CNO isotopes from \citet{Lodders2009}, one obtains:
\begin{equation}
\label{eq:epsHsolar}
    \epsilon_\mathrm{H} = 5.93 \times 10^{13}\, \frac{Z_{\mathrm{CNO, scaled\,\odot}}}{0.01}\mathrm{\quad erg\,g^{-1}\,s^{-1}}\,
\end{equation}
which is remarkably similar to the value using the $\beta$-limited $Z$ value reference, though only by coincidence, not by design.  

Throughout the remainder of this paper all values of metallicity refer exclusively to the $\beta$-limited CNO metallicity, which is the relevant value for {\sc settle}.  The reference value of $0.01$ used here is indeed very close to the solar mass fraction of CNO isotopes given by \citet{Lodders2009} as $0.01001$. 

Summarising, the time to burn all of the hydrogen ($t_{\mathrm{H}}$) is dependent only on the metallicity and initial hydrogen abundance. The time to burn hydrogen is thus

\begin{equation}
    \label{eq:Htime}
    t_{\mathrm{H}} = \frac{Y_{\mathrm{^1H}}}{Y_{\mathrm{CNO}}}\frac{t_{\mathrm{\frac{1}{2}, ^{15}O}} + t_{\mathrm{\frac{1}{2}, ^{14}O}}}{4\ln{2}}\, = 20.45 \left(\frac{X_0}{0.7}\right)\left(\frac{0.01}{Z_{\mathrm{CNO}}}\right)\,\mathrm{h}
\end{equation}
where $X_0$ is the initial hydrogen mass fraction of the accreted fuel.  This sets the minimum recurrence time required for pure helium bursts.  Note that we have updated this equation from the original work of \citet{Cumming2000}, and it is the time to burn hydrogen locally on the neutron star surface, for the observer this timescale is longer by a redshift factor $1+z$.
For higher accretion rates ($\dot{m}\gtrsim0.04\,\dot{m}_{\mathrm{Edd}}$\footnote{$\dot m_{\mathrm{Edd}}$ is the Eddington accretion rate, given by $\dot{m}_{\mathrm{Edd}}=1.75\frac{0.7}{1+X}\times10^{-8}\,\mathrm{M}_{\odot}\,\mathrm{yr}^{-1}$, where $X$ is the H mass fraction}) helium burning becomes unstable before sufficient time has passed for all of the hydrogen to be burned, resulting in ignition occurring in a hydrogen rich environment. For lower ($\dot{m}\lesssim0.04\,\dot{m}_{\mathrm{Edd}}$) accretion rates, there is sufficient time for all of the hydrogen to be burned before ignition, allowing a pure helium layer to accumulate before ignition, and a helium rich environment for the burst. 

The column depth at which hydrogen runs out ($y_\mathrm{d}$) is given by:

\begin{equation}
    y_{\mathrm{d}} = 6.9 \times 10^8 \left(\frac{\dot m}{0.01 \dot m_{Edd}}\right) \left(\frac{0.01}{Z_{\mathrm{CNO}}}\right)\left(\frac{X_{\mathrm{0}}}{0.7}\right)\quad \mathrm{ g\,cm^{-2}}
\end{equation}

If helium ignites at a column depth $y<y_\mathrm{d}$, a mixed H/He burst occurs; and if helium ignites at a column depth $y>y_\mathrm{d}$, a pure helium layer has time to accumulate and ignition occurs in a pure helium environment. 

Once ignition conditions are met, the burst parameters are calculated based on the total mass in the ignition column and its composition. The burst energy is calculated by assuming complete burning of the H/He layer, and the accreted material is assumed to cover the whole surface of the neutron star. The thickness of the accumulated layer at ignition determines the recurrence time of the bursts.

\subsubsection{Updates to {\sc settle}}

When hydrogen remains at the time of ignition, and bursts ignite in a mixed H/He environment, there is extra energy released by the triple-alpha process due to proton capture by $^{12}$C. To avoid a discontinuity in accretion rate--recurrence time parameter space at the transition zone between mixed H/He bursts and pure He bursts, we gradually switched on the factor that multiplies the triple alpha reaction energy generation rate to account for the $^{12}\mathrm{C}(\mathrm{p},\gamma)$ reaction when protons begin running out at the critical H fraction of $X=1/7$ (since the ratio of protons to alpha particles is $2:3$). The gradual switch on of this multiplicative factor was not a feature of \citet{Cumming2000}'s original code. 

We updated some of the energy and rate values used in {\sc settle}. We corrected the energy generation of hydrogen burning via the CNO cycle such that $\epsilon_\mathrm{H} = 5.94\times10^{15}\,\mathrm{erg}\,\mathrm{g}^{-1}\,\mathrm{s}^{-1}$ (used to be $5.8\times10^{15}\,\mathrm{erg}\,\mathrm{g}^{-1}\,\mathrm{s}^{-1}$). We also used an updated version for the total nuclear energy released per unit mass in an X-ray burst as calibrated by \citet{Goodwin2018} using \kepler models, $Q_{\mathrm{nuc}} = 1.35 + 6.05\,\bar{X}\,\mathrm{MeV}/\mathrm{nucleon}$, where $\bar{X}$ is the average hydrogen mass fraction of the ignition column. The ignition depth correction we applied due to differences between {\sc settle} and \kepler is described in Section \ref{sec:comparison}

Using \kepler models with fully time-dependent accretion rates, \citet{zacpaper} found that using constant accretion rates (averaged between bursts) can systematically underestimate the burst recurrence times when the slope of the accretion rate is negative (see Figure 4 of that paper). Because {\sc settle} implicitly uses averaged accretion rates, we accounted for this effect by multiplying each predicted recurrence time by a correction factor, as determined by the linear approximation from \citet{zacpaper}. These correction factors ranged from approximately $0.9$--$1.0$ for our modelled bursts. The correction factors we adopted are specific to SAX J1808.4--3658 and will either need to be recalculated or made more generally applicable if this method is applied to a different system. To rigorously test the influence of these correction factors we ran the code without the scaling factors and found that while the central value of the distance, hydrogen fraction, metallicity, and base flux shifted, the parameters still agreed within 1-$\sigma$ uncertainty of the model run with the scaling factors.

\subsection{Code Verification}\label{sec:comparison}

A simple code such as {\sc settle} can be run in seconds, enabling multiple instances of the code to be run, and a marginalising method such as MCMC possible. Other codes, such as \kepler \citep{woosley2004}, use bigger nuclear reaction networks and a more complex treatment of the neutron star atmosphere during outburst, at the cost of computational expense and runtime. \kepler solves the full time-dependent stellar structure equations, following the evolution of temperature and composition at each depth in the neutron star envelope over many cycles of bursts. To assess the limitations of {\sc settle}, we ran a comparison with some \kepler models, and implemented corrections to {\sc settle} to account for some of the differences we found. \citet{woosley2004} provided a limited comparison between some {\sc settle} and \kepler models, and found that, in general, the agreement was good, although with {\sc settle} consistently overestimating the burst energies and recurrence times. The authors attribute the discrepancy between the {\sc settle} and \kepler models to the lack of compositional inertia in the {\sc settle} models.

We compared the predicted fluences ($E_{\mathrm{b}}$) and recurrence times ($\Delta t$) of the two codes for a range of hydrogen mass fractions ($X$), CNO mass fractions ($Z$), accretion rates ($\dot{m}$) and base fluxes ($Q_{\mathrm{b}}$). In order to allow meaningful comparison, we redshifted the \kepler recurrence times by multiplying by $1+z$, as \kepler calculates recurrence time in the reference frame of the neutron star, whereas {\sc settle} calculates recurrence time in the reference frame of a distant observer. In both codes we fixed $M_{\mathrm{NS}}=1.4\,\mathrm{M}_{\odot}$ and $R_{\mathrm{NS}}=11.2\,\mathrm{km}$.  We also increased the {\sc settle} $Q_{\mathrm{b}}$ value by $0.1\,\mathrm{MeV}/\mathrm{nucleon}$ when comparing to \kepler to account for thermal inertia \citep{Cumming2000}. 

\begin{figure}
	\includegraphics[width=\columnwidth,viewport=8 0 396 298]{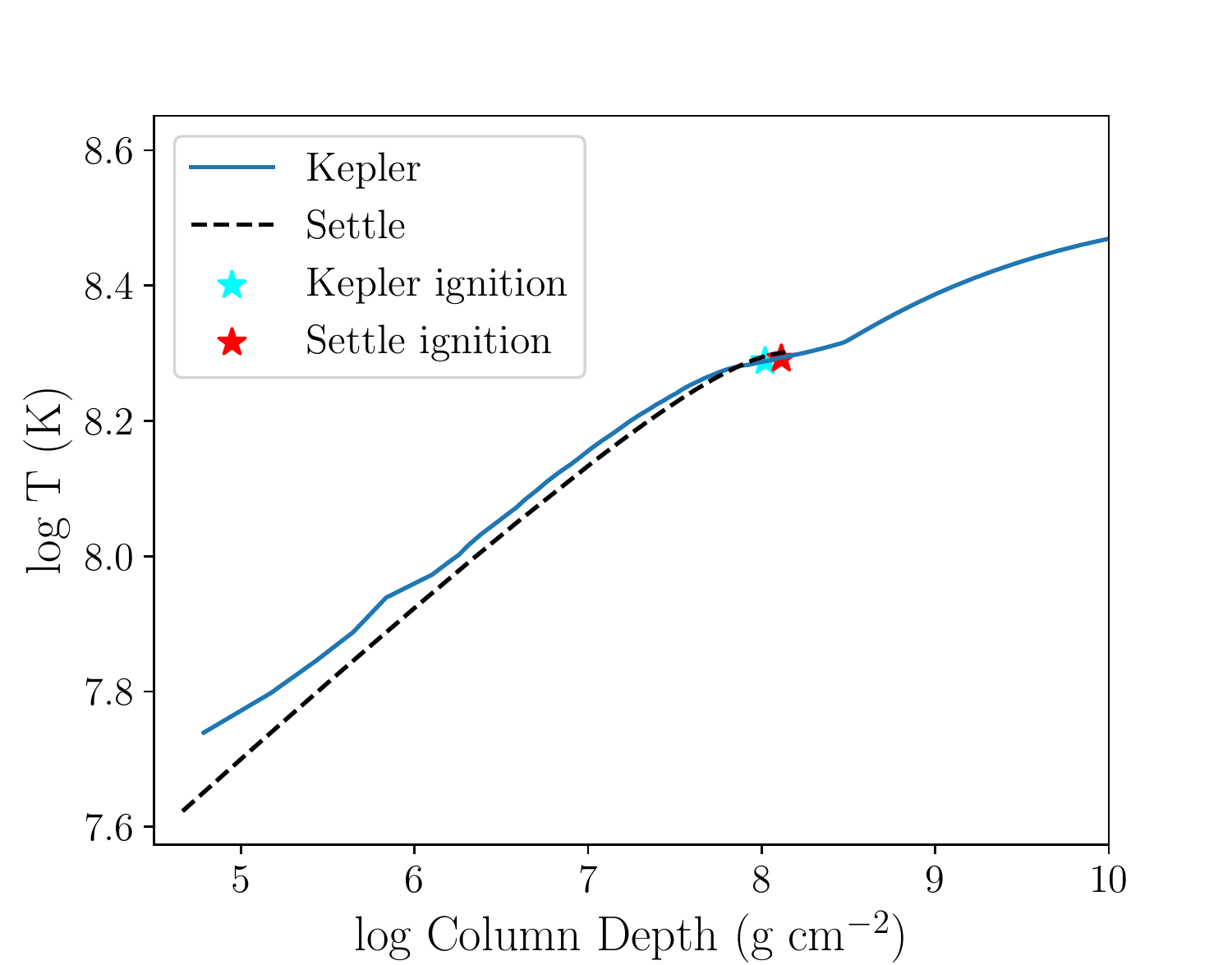}
    \caption{Comparison of the temperature profile of {\sc settle} (black dashed) and \kepler (blue solid) models at the ignition of a burst. Stars indicate the ignition depth of the burst for each model. Model conditions: $X=0.2$, $z=0.02$, $\dot{m}=0.2\,\dot{m}_{\mathrm{Edd}}$, $Q_{\mathrm{b}}=0.1\,\mathrm{Mev}/\mathrm{nucleon}$.
    }
    \label{fig:tempprofile}
\end{figure}

On initial comparison we found the {\sc settle} recurrence times and fluences were at least $\sim$ 30$\%$ higher than the \kepler predictions. We compared the temperature profiles of the ignition column for each code, shown in Figure \ref{fig:tempprofile}.

\begin{figure}
	\includegraphics[width=\columnwidth,viewport=25 11 630 355]{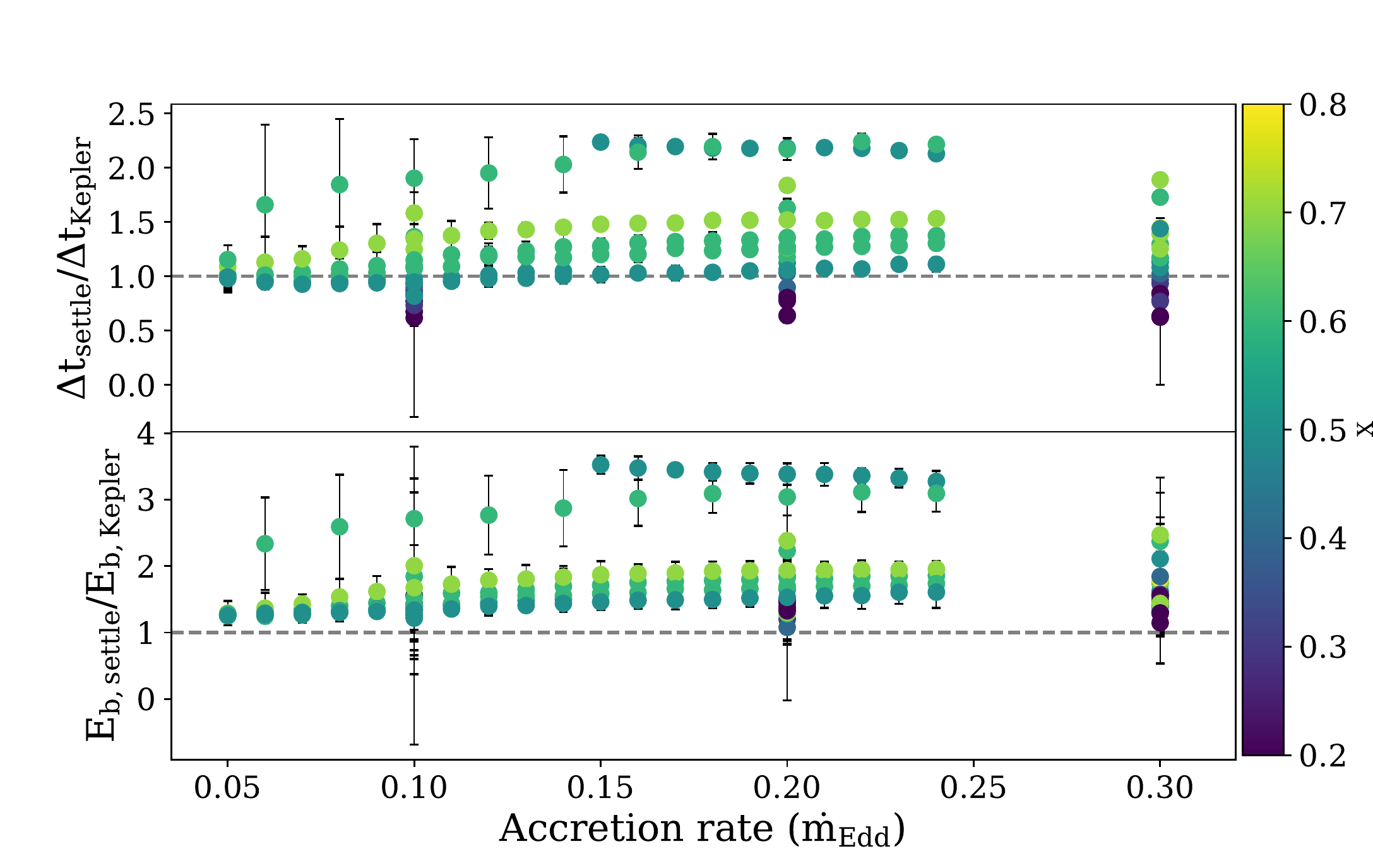}
	\caption{Comparison of the predicted recurrence times (top) and burst fluences (bottom) for the {\sc settle} and \kepler models for a range of accretion rates ($\dot{m}$, expressed as a fraction of $\dot{m}_{\mathrm{Edd}}$), after the ignition depth in {\sc settle} has been reduced by $45\%$. The dashed line shows the one-one relation between the two predictions. Some of the models have different CNO metallicities and base heating, causing some variances in the predictions for the same accretion rate and hydrogen mass fraction.}
    \label{fig:settleKepcomparison}
\end{figure}

We found the temperature profile of {\sc settle} was overall cooler than \kepler, causing a longer recurrence time, deeper ignition, and thus brighter bursts. Since \kepler provides a more complex analysis of the underlying physics, we decided to correct the ignition depth of {\sc settle} such that it matched the \kepler ignition depth. To do this we multiplied the {\sc settle} ignition depth by a factor of $0.65$. Reducing the ignition depth reduced the recurrence time and fluences of the {\sc settle} bursts and produced the comparison shown in Figure \ref{fig:settleKepcomparison}. 

In Figure \ref{fig:settleKepcomparison} for higher hydrogen fractions ($X = 0.6$ and $0.7$) there is a significant difference between the predicted fluences and recurrence times of the {\sc settle} and \kepler bursts across all accretion rates. There is much better agreement at lower hydrogen fractions ($X < 0.5$) and lower accretion rates ($\dot{m}<0.1\,\dot{m}_{\mathrm{Edd}}$ between the two codes. For the purpose of this paper we note that for the source we are investigating, \texttt{SAX J1808.4--3658}, observations indicate that is has a slightly depleted hydrogen fraction and low ($\approx4\%$ Eddington) accretion rate, so it is within the parameter space that {\sc settle} agrees well with \kepler. 

Fundamental differences between the two models can be attributed primarily to the fact that {\sc settle} has a one-zone ignition criterion which uses a local approximation for the cooling, whereas \kepler follows the detailed cooling as a function of depth. Other differences could arise due to the fact that {\sc settle} assumes there is no triple-alpha burning during the accumulation of fuel, {\sc settle} assumes that all of the accreted fuel is burnt in a single burst, and the burst train is generated without accounting for subsequent bursts occurring in the ashes of previous bursts. Finally, {\sc settles} uses an approximate formula for the nuclear energy generation of the bursts, that has been calibrated to \kepler predictions, but still is not the same as calculating the full nuclear reaction network, with thermal inertia included. As a caveat we note that calibrating to {\sc kepler} introduces a different kind of model bias. Since there have been no comprehensive studies done between {\sc kepler} and other similar codes, we do not understand how accurately {\sc kepler} models the true physics of these systems.

\subsection{MCMC}\label{sec:mcmc}

We used a Python implementation of MCMC \citep[{\sc emcee};][]{Foreman2013} to estimate posterior distributions of the properties predicted by {\sc settle}, based on the known values of the observed properties of the burst train (time, $E_{\mathrm{b}}$ and $\alpha$). We defined a likelihood function, given by Equation \ref{likelihood}, and prior ranges for each parameter. We used a simple Gaussian likelihood function where the variance was underestimated by some fractional amount, $f$, and the overall likelihood is the sum of the individual likelihoods for each parameter $i$,

\begin{equation}
\label{likelihood}
\centering
    L(x_i|\theta, \mathrm{model}) = \frac{1}{\sigma_i f_i \sqrt{2\pi}} \exp\left[\frac{-(x_i - \mathrm{model})^2}{2 f_i \sigma_i}\right]
\end{equation}

where $\sigma_i$ is the observed uncertainty, $x_i$ is the observed parameter value (time, fluence, and alpha), model is the parameter value predicted by the model, and $\theta$ are our parameters of interest. The parameters of interest we defined are hydrogen mass fraction $X$, metallicity $Z$, which is defined as the mass fraction of CNO elements (all in $^{14}$O and $^{15}$O), base flux $Q_{\mathrm{b}}$, neutron star mass $M$, neutron star radius $R$ and three scaling factors, $r_{\mathrm{1}}$, $r_{\mathrm{2}}$, and $r_{\mathrm{3}}$.

The 3 scaling factors, $r_{\mathrm{1}}$, $r_{\mathrm{2}}$, and $r_{\mathrm{3}}$ are defined as in \citet{galloway2006}, given in Equations \ref{eq:r1}, \ref{eq:r2}, and \ref{eq:r3},

\begin{equation}
    r_{\mathrm{1}} = \frac{\dot m_i}{c_{\mathrm{bol}}F_{\mathrm{p},i}\xi_{\mathrm{p}}}
    \label{eq:r1}
\end{equation}

where $\dot m_i$ is the accretion rate between burst $i-1$ and burst $i$, $\xi_{\mathrm{p}}$ is the persistent anisotropy (defined in Section \ref{sec:scaling}, $F_{\mathrm{p},i}$ is the mean $2-25\,\mathrm{keV}$ persistent flux between burst $i-1$ and $i$ (in units of $10^{-9}\,\mathrm{erg}\,\mathrm{cm}^{-2}\,\mathrm{s}^{-1}$, from interpolation) and $c_{\mathrm{bol}}$ is the bolometric flux correction.  

\begin{equation}
    r_{\mathrm{2}} = \frac{\alpha}{\alpha_{\mathrm{pred}}} 
    \label{eq:r2}
\end{equation}

where $\alpha$ is the observed value (Equation \ref{eq:alpha}) and $\alpha_{\mathrm{pred}} = 290/Q_{\mathrm{nuc,pred}}$.

\begin{equation}
    r_{\mathrm{3}} = \frac{E_{\mathrm{b}}/10^{-9}}{L_{\mathrm{b,pred}}\,\xi_{\mathrm{b}}} 
    \label{eq:r3}
\end{equation}
where $L_{\mathrm{b,pred}}$ is the predicted burst luminosity in units of $10^{39}\,\mathrm{erg}\,\mathrm{s}^{-1}$ and $E_{\mathrm{b}}$ is the observed burst fluence. 

We initialised the parameters in a tiny ($\sim10^{-3}$) Gaussian ball around the initial value. We ran the MCMC chains with 300 walkers for 2,000 steps, and discarded the first 200 steps in each chain to account for burn-in. See Section \ref{sec:chainconvergence} for information on the tests we ran to ensure the MCMC chains were fully converged and independent of initial walker positions.

\subsection{Priors}
\begin{figure}
    \centering
    \includegraphics[width=\columnwidth,viewport=12 0 389 255,clip]{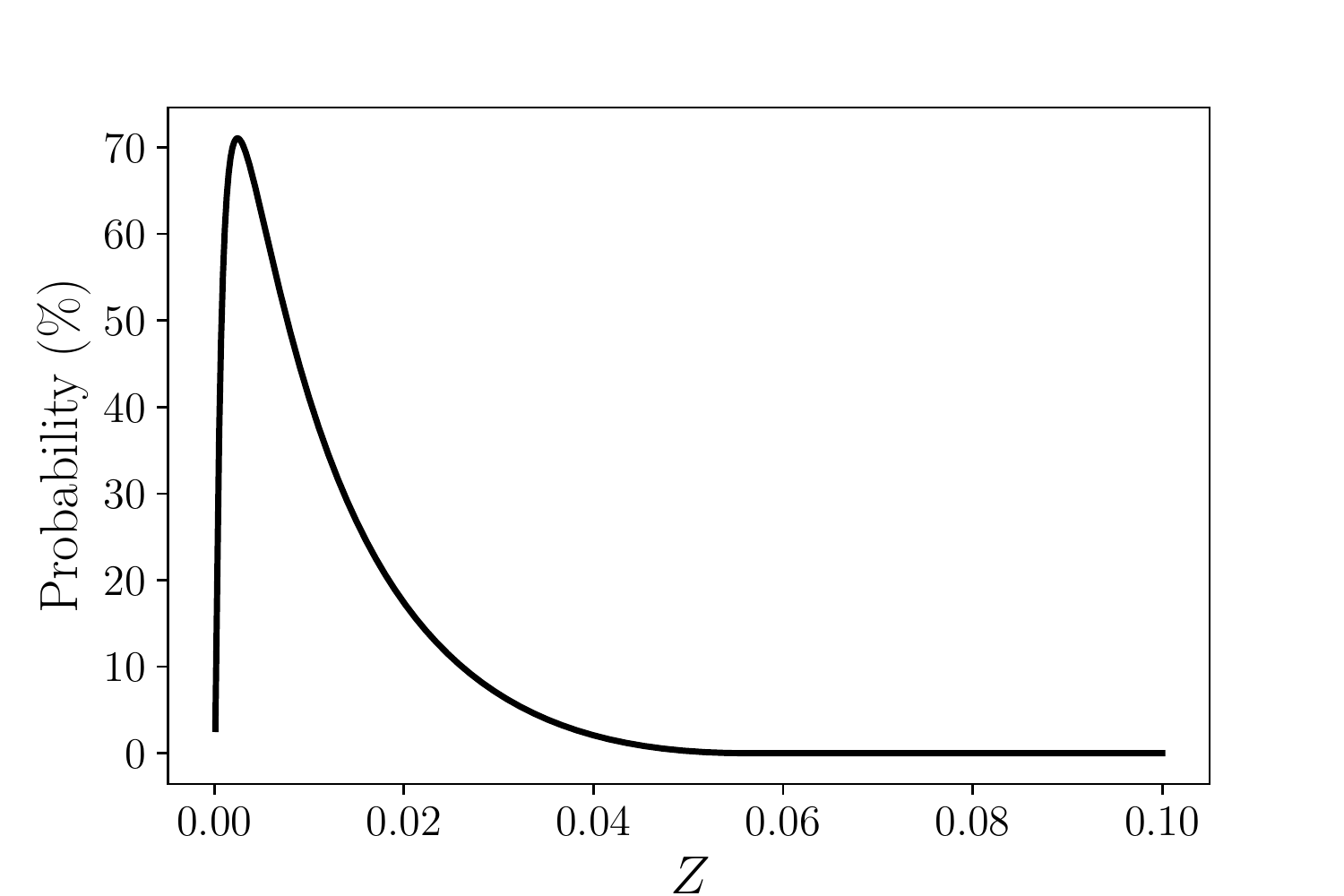}
    \caption{The probability distribution assumed for metallicity ($Z$), based on a simple model of the Galaxy in the location of \texttt{SAX J1808.4--3658} at $2$--$4\,\mathrm{kpc}$.}
    \label{fig:Zprior}
\end{figure}

For each parameter except metallicity, mass and radius, we used a uniform prior, where we chose an acceptable range based on observations or theoretical assumptions. For metallicity ($Z_{\mathrm{CNO}}$) we chose a more informed prior, using a synthetic survey of the Milky Way in the location of \texttt{SAX J1808.4--3658} at $2$--$4\,\mathrm{kpc}$ generated by the Galaxia code \citep{Sharma2011} to determine a likely distribution of the metallicity, given by Equation \ref{eq:metallicityprior} and shown in Figure \ref{fig:Zprior}.

\begin{equation}
\label{eq:metallicityprior}
  \mathrm{[Fe/H]} \sim \phi (3 + \frac{3}{4}) - 3
\end{equation}

where $\phi \sim \mathcal{B}(10,3)$ is a Beta distribution with $\alpha = 10$ and $\beta = 3$, and the
linear translation limits the metallicity prior to have support between $[\textrm{Fe/H}] = -3$ and $+0.75$ (i.e., $0.00001 < Z_{\mathrm{CNO}} < 0.056$).

\begin{figure}
\centering
	\includegraphics[width=0.8\columnwidth,viewport=0 4 363 363]{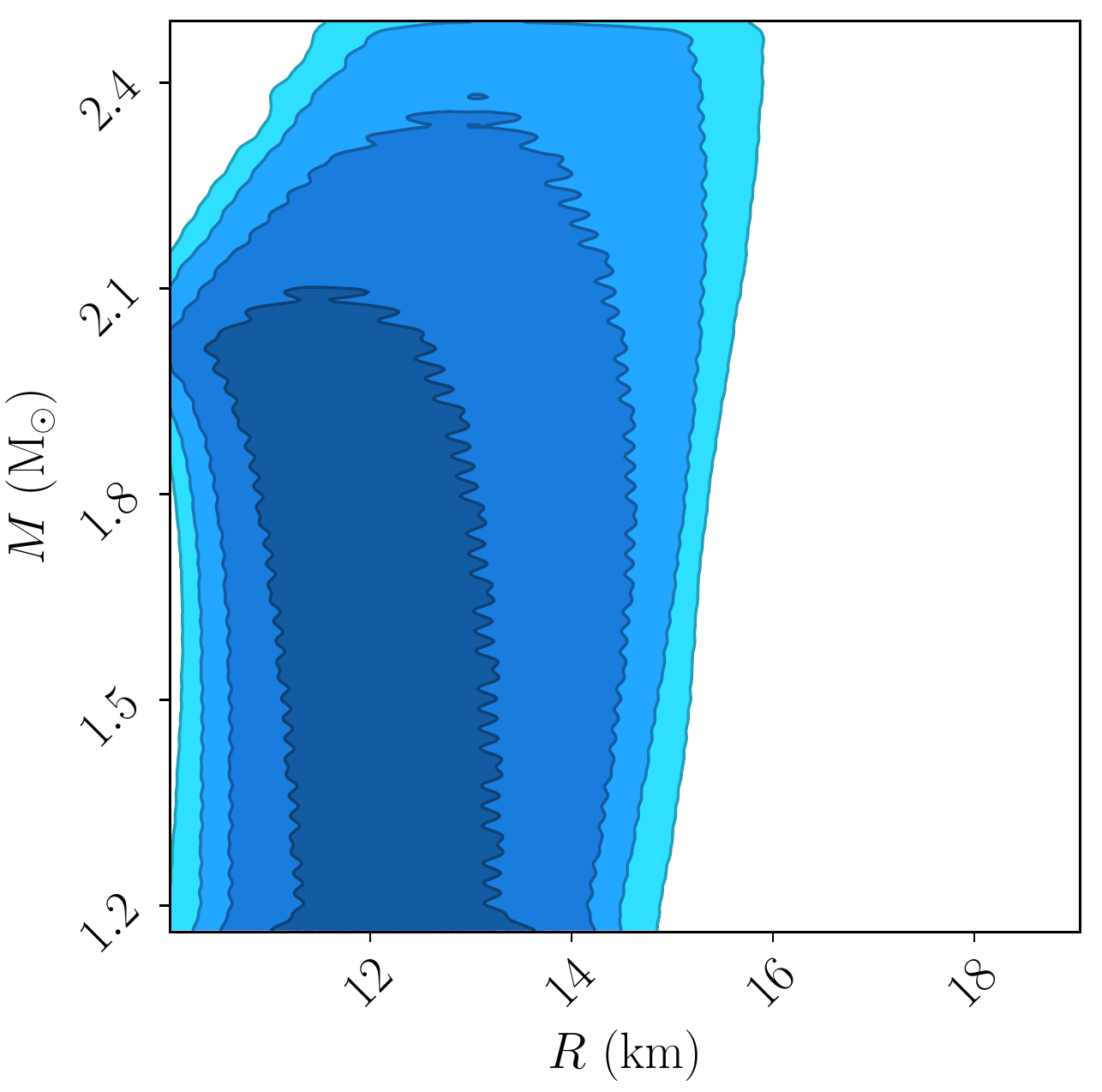}
	\label{fig:mr_prior}
    \caption{\citet{Steiner2018} probability distribution for NS radius as a function of mass based on observations of low mass X-ray binaries in globular clusters. Contour levels shown are 1$\sigma$ intervals up to 5$\sigma$.}
\end{figure}

For mass and radius we also imposed a more informed prior based on the models of \citet{Steiner2018}, who used observations of eight quiescent low mass X-ray binaries in globular clusters to determine the neutron star mass--radius curve and the equation of state. \citet{Steiner2018} used a fixed mass grid and varied radius according to their chosen equation of state and observations of radii. We chose the simplest of their models (`baseline') and used the probability distribution given in Figure \ref{fig:mr_prior} for the relationship between mass and radius. We also constrained mass and radius to: $1.15\,\mathrm{M_{\odot}} < M < 2.5\,\mathrm{M_{\odot}}$ and $9\,\mathrm{km} < R < 17\,\mathrm{km}$. This prior effectively checks if the masses and radii predicted by the model are consistent with the expected mass-radius relation of low mass X-ray binaries in globular clusters. It provides a probability of the predicted radius given the predicted mass. 

\section{Results}
\label{sec:results}
We report the posterior limits for the predicted system parameters in Table \ref{tab:1808_parameters}, reporting the mean value and the 68th percentile limits of the distributions for the uncertainties. Additional model runs are shown in the Appendix, including one that replicates \citet{galloway2006} with fixed mass and radius and one with a uniform prior in mass and radius.
{
\renewcommand{\arraystretch}{1.6}
\begin{table}
    
	\centering
	\caption{\texttt{SAX J1808.4--3658} derived neutron star parameters}
	\label{tab:1808_parameters}
	\begin{tabular}{lcc}
		\hline
		Parameter & Value & Units\\
		\hline
		X & $0.58^{+0.13}_{-0.14}$ & \\
		Z & $0.013^{+0.006}_{-0.004}$& \\
		$Q_{\mathrm{b}}$ & $0.4^{+0.3}_{-0.2}$& $\mathrm{MeV/nucleon}$\\
		$M$ & $1.5^{+0.6}_{-0.3}$ & $\mathrm{M}_{\odot}$ \\
		$R$ & $11.8^{+1.3}_{-1.0}$ & $\mathrm{km}$ \\
		$\dot m_{\mathrm{max}}$ &$0.037^{+0.002}_{-0.002}$ & $\dot m_{\mathrm{Edd}}$ \\
		$g$ & $1.9^{+0.7}_{-0.4}$&$10^{14}\,\mathrm{cm}\,\mathrm{s}^{-2}$ \\
		$1+z$ & $1.27^{+0.13}_{-0.05}$ & \\
		$d$ & $3.3^{+0.3}_{-0.2}$ & $\mathrm{kpc}$ \\
		$\xi_{\mathrm{b}}$ & $0.74^{+0.10}_{-0.10}$ \\
		$\xi_{\mathrm{p}}$ & $0.87^{+0.12}_{-0.10}$ \\
		$\cos i$ & $0.36^{+0.07}_{-0.04}$\\
		\hline
	\end{tabular}
	\small{\\\textbf{Notes:}$g$ and $1+z$ are calculated based on the $M$ and $R$ values, using Equation \ref{eq:grav}. $\dot{m}_{\mathrm{max}}$ is the maximum inferred accretion rate using Equation \ref{eq:fptomdot}.}
\end{table}
}

\begin{figure*}
	\includegraphics[width=0.67\textwidth]{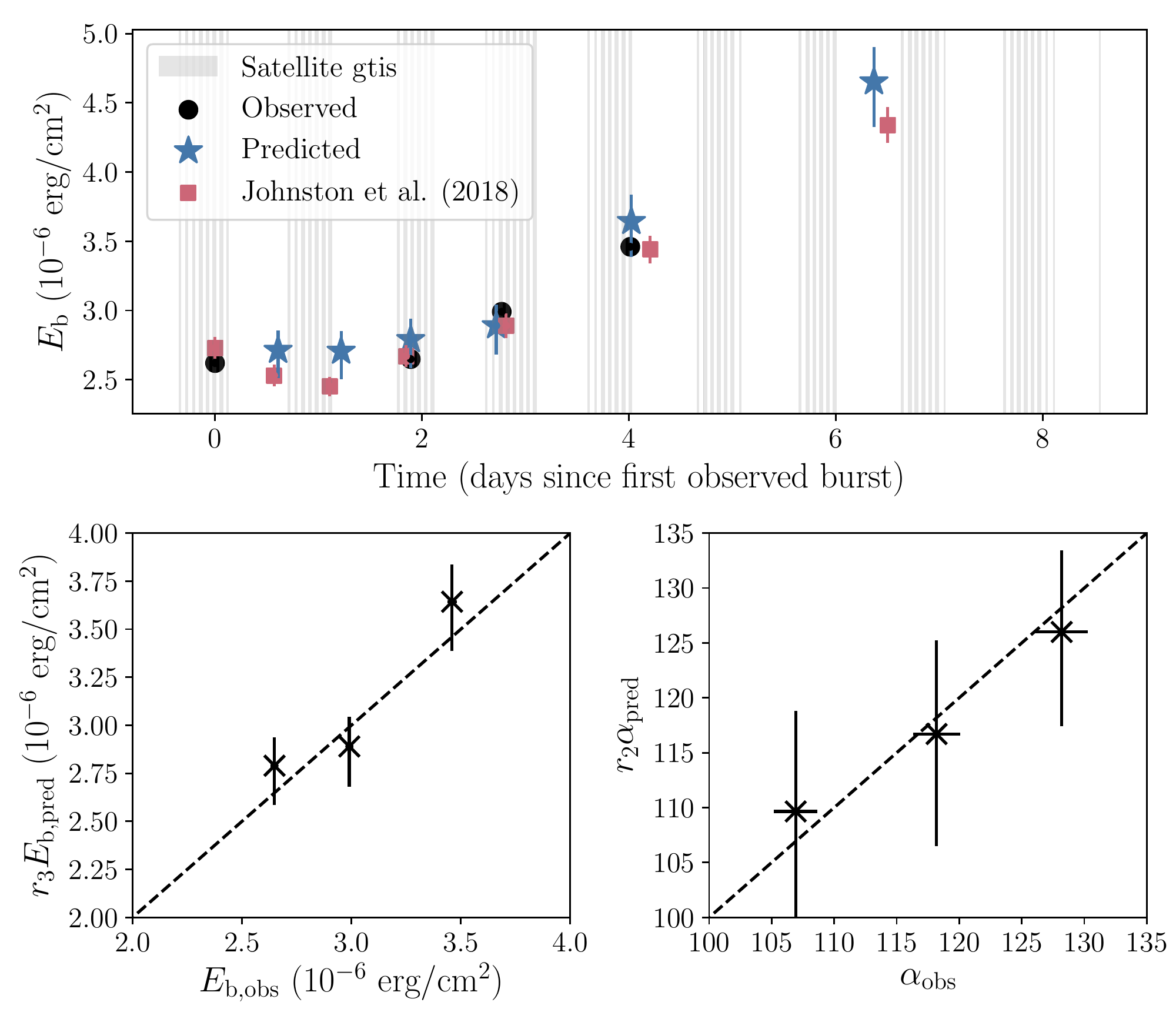}
    \caption{Top: predicted burst times and fluences (blue stars), observed bursts times and fluences (black circles) and predicted burst times and fluences from \citet{zacpaper} (red squares). The grey regions show the satellite observing periods (`good time intervals'). Bottom left: $r_3$ times the predicted burst fluences vs observed burst fluences. Dashed line is the one-to-one relation. Bottom right: $r_2$ times the predicted alphas vs observed alphas. Dashed line is the one-to-one relation. }
    \label{fig:matchedbursts}
\end{figure*}

The times and fluences of the 6 predicted bursts are plotted in Figure \ref{fig:matchedbursts}, as well as the 4 observed bursts. We predict 3 bursts that were not observed, but that cannot be ruled out by observations, since they occur during periods that the telescope was not observing. The two bursts predicted between the first and second observed bursts were also predicted by \citet{galloway2006} and \citet{zacpaper}. The predicted burst properties agree reasonably well with the predictions from \citet{zacpaper}, who used \kepler to model this outburst without doing an exhaustive parameter space exploration such as we have done in this work. Interestingly, \citet{zacpaper} also predict a very bright burst at late time, $\approx 7$ days after the first observed burst. 

Burst properties for individual bursts inferred by the model are listed in Table \ref{tab:inferredbursts}. The recurrence time for each burst is predicted to fall in the range 14 -- 56 hours. The time to exhaust hydrogen and ignite bursts in a pure helium environment is 13.0~hours at this accreted fuel composition (Equation \ref{eq:Htime}). These predicted recurrence times and their respective accretion rates indicate that sufficient time passes between each burst for the accreted hydrogen to be completely burned by the hot CNO cycle, and we have ignition occurring in a pure helium environment, as has been previously concluded by \citet{galloway2006, zacpaper}. The low $\bar{X}$ values for each burst confirms this, with $\bar{X} < 0.2$ for all bursts.
{
\renewcommand{\arraystretch}{1.6}
\begin{table}
	\centering
	\caption{\texttt{SAX J1808.4--3658} predicted burst parameters}
	\label{tab:inferredbursts}
	\begin{tabular}{lcccr}
		\hline
		Burst & $\bar{X}$ & $\alpha$ & $\Delta t$ (hours) \\
		\hline
		2 & $0.20^{+0.06}_{-0.06}$ & $108^{+12}_{-9}$& $14.7$\\
		3 & $0.20^{+0.06}_{-0.06}$& $107^{+12}_{-9}$ & $14.4$\\
		4*  & $0.19^{+0.05}_{-0.06}$& 110$^{+11}_{-9}$ & 16.1\\
		5*  & $0.16^{+0.04}_{-0.04}$& $117^{+10}_{-9}$ & $19.4$ \\
		6* & $0.12^{+0.03}_{-0.03}$ & $126^{+9}_{-7}$ & $31.2$ \\
		7  & $0.07^{+0.02}_{-0.02}$ & $139^{+6}_{-5}$ & $56$ \\

		\hline
	\end{tabular}
	\\
	  \small{\textbf{Notes:} "*" indicates this burst was observed by \textit{RXTE}.  Burst 1 is excluded as it was used as a reference burst and we did not infer properties for this burst. $\bar{X}$ is the average hydrogen mass fraction of the ignition column.}
	  
\end{table}
}

\begin{figure*}
	\includegraphics[width=0.75\textwidth,viewport=8  32 685 685,clip]{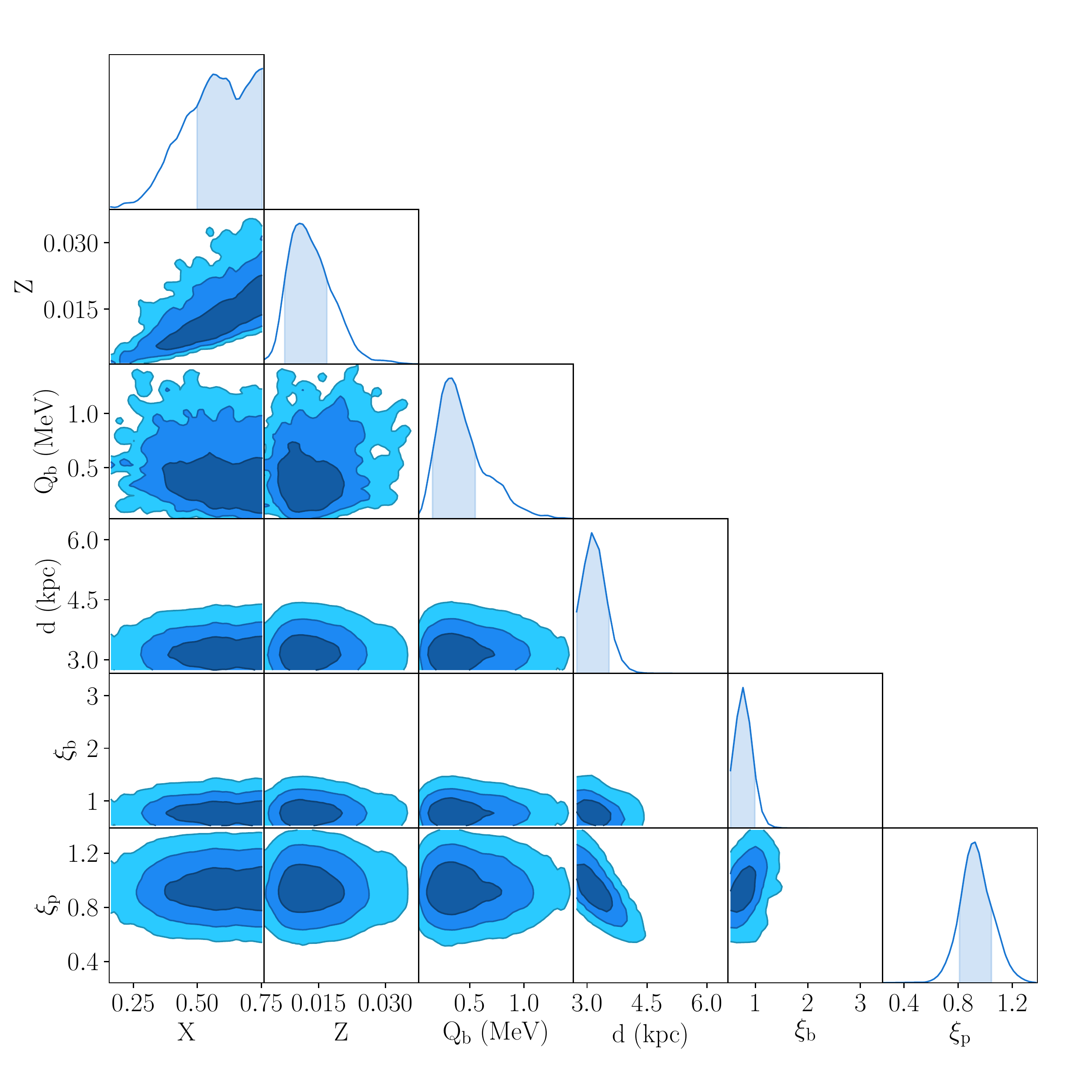}
    \caption{2-dimensional projection of the posterior probabilities of hydrogen mass fraction ($X$), metallicity ($Z$), base heating rate ($Q_{\mathrm{b}}$), distance ($d$) and anisotropy factors ($\xi_{\mathrm{b}}$ and $\xi_{\mathrm{p}}$) of \texttt{SAX J1808.4--3658}. Histograms along the diagonal show the marginalised probabilities of the individual parameters. Contour levels are the 1, 2, and 3 $\sigma$ bounds.}
    \label{fig:mrxzcorrelation}
\end{figure*}

The probability contours and posterior distributions for $X$, $Z$, $Q_{\mathrm{b}}$, $d$, $\xi_{\mathrm{b}}$, and $\xi_{\mathrm{p}}$ are plotted in Figure \ref{fig:mrxzcorrelation}. There is a positive correlation between $X$ and $Z$, which was also found by \citet{galloway2006}. This correlation is due to the relationship between recurrence time and fuel composition, since provided all of the H is exhausted before ignition, there are multiple combinations of $X$ and $Z$ that reproduce the observed recurrence time.

\subsection{Distance and Anisotropies}
\label{sec:scaling}

We used the scaling parameters $r_{\mathrm{1}}$, $r_{\mathrm{2}}$, and $r_{\mathrm{3}}$ to obtain posterior distributions for the distance and emission anisotropies of the source, and thus the inclination, shown in Figure \ref{fig:mrxzcorrelation}. Relating the scaling parameters to their corresponding physical parameters gives: 

\begin{equation}
    r_{\mathrm{2}} = 0.74816 \left(\frac{M_{\mathrm{NS}}}{1.4\,\mathrm{M_{\odot}}}\right)
    \left(\frac{R_{\mathrm{NS}}}{11.2\,\mathrm{km}}\right)^{-1}
    \left(\frac{Q_{\mathrm{nuc}}}{Q_{\mathrm{nuc,pred}}}\right)^{-1}
    \left(\frac{\xi_{\mathrm{p}}}{\xi_{\mathrm{b}}}\right)
\end{equation}

\begin{equation}
    r_{\mathrm{3}} = 63.23 \left(\frac{d}{10\,\mathrm{kpc}}\,\right)^{-2}
    \left(\frac{R_{\mathrm{NS}}}{11.2\,\mathrm{km}}\right)^{2}
    \left(\frac{1+z}{1.258}\right)^{-1}
    \left(\frac{Q_{\mathrm{nuc}}}{Q_{\mathrm{nuc,pred}}}\right) 
    \xi_{\mathrm{b}}
\end{equation}

Assuming $Q_{\mathrm{nuc}} = Q_{\mathrm{nuc, pred}}$, this gives the following set of equations which enable calculation of d, $\xi_{\mathrm{b}}$ and $\xi_{\mathrm{p}}$ using the three scaling factors.

\begin{equation}
    \xi_{\mathrm{p}} = r_{\mathrm{1}} \left(\frac{10}{d}\right)^2
\end{equation}
\begin{equation}
     \xi_{\mathrm{b}} = \frac{r_{\mathrm{2}} \xi_{\mathrm{p}}}{0.74816}
\end{equation}
\begin{equation}    
    d = 10\sqrt{\frac{63.23 \xi_{\mathrm{b}}}{r_{\mathrm{3}}}}\quad \mathrm{kpc}
\end{equation}

Using these equations, we obtained posteriors on distance ($d$), burst anisotropy ($\xi_{\mathrm{b}}$) and persistent anisotropy ($\xi_{\mathrm{p}}$), shown in Figure \ref{fig:mrxzcorrelation} and values reported in Table \ref{tab:1808_parameters}. \citet{he2016} modelled the relationship between anisotropy factors and inclination in low mass binaries with different disc shapes. They considered 4 different disc geometries, motivated by an effort to match observations of high reflection fractions observed in a couple of superbursts. Since \texttt{SAX J1808.4--3658} has never shown evidence of such high reflection fractions, we chose the simplest disc geometry described by \citet{he2016}, Model~A, corresponding to a flat disc. \citet{he2016} Model~A is approximated by Equation \ref{eq:fujimoto}, which we used to also obtain a posterior distribution for cos $i$, based on the inferred anisotropy factors.

\begin{equation}
\label{eq:fujimoto}
    \frac{\xi_{\mathrm{p}}}{\xi_{\mathrm{b}}} = \frac{0.5 + |\mathrm{cos} i|}{2|\mathrm{cos} i|}
\end{equation}

where $i$ is the inclination of the system, the angle between the rotation axis and the direction from which the system is viewed by an observer. 

For this disc Model~A, we infer a distance to \texttt{SAX J1808.4--3658} of $3.3^{+0.3}_{-0.2}\,\mathrm{kpc}$, which agrees within uncertainty of the predictions of \citet{galloway2006} of 3.1--3.8\,kpc. We note that \citet{galloway2006} did not account for the anisotropy of emission when inferring their distance estimate. Applying our predicted anisotropy factors to their distance estimate reduces their estimate to 2.7--3\,kpc, still within our predicted distance range.

\subsection{Neutron Star Mass and Radius}

We infer a neutron star mass of $M_{\mathrm{NS}}=1.5^{+0.6}_{-0.3}\,\mathrm{M}_{\odot}$ based on the gravity and gravitational redshift required by the model that gives the best match to the observations. Since the neutron star in this system has been accreting matter periodically, we expect it to be more massive than a single non-binary neutron star \citep[e.g.][]{Kiziltan2013}. Canonically, the neutron star mass is assumed to be $M_{\mathrm{NS}}=1.4\,\mathrm{M}_{\odot}$, and recent works constraining NS mass and radii have found the NS mass could be between $1.1$--$2.2\,\mathrm{M}_{\odot}$ \citep[e.g.,][]{Antoniadis2016}. There has been extensive work in trying to estimate the mass of \texttt{SAX J1808.4--3658} with no consensus on the value, but the general finding that it is more massive than the canonical value of 1.4\,$\mathrm{M}_{\odot}$ \citep[e.g.][]{Heinke2009,Elebert2009,Morsink2011}. If the NS mass is larger than the usually assumed NS mass, this could indicate that some of the accreted matter is not burned in the observed X-ray bursts, as was modelled by \citet{Goodwin2018}, however, due to the uncertainty of our mass prediction this result is not significant. 

\begin{figure}
	\includegraphics[width=\columnwidth]{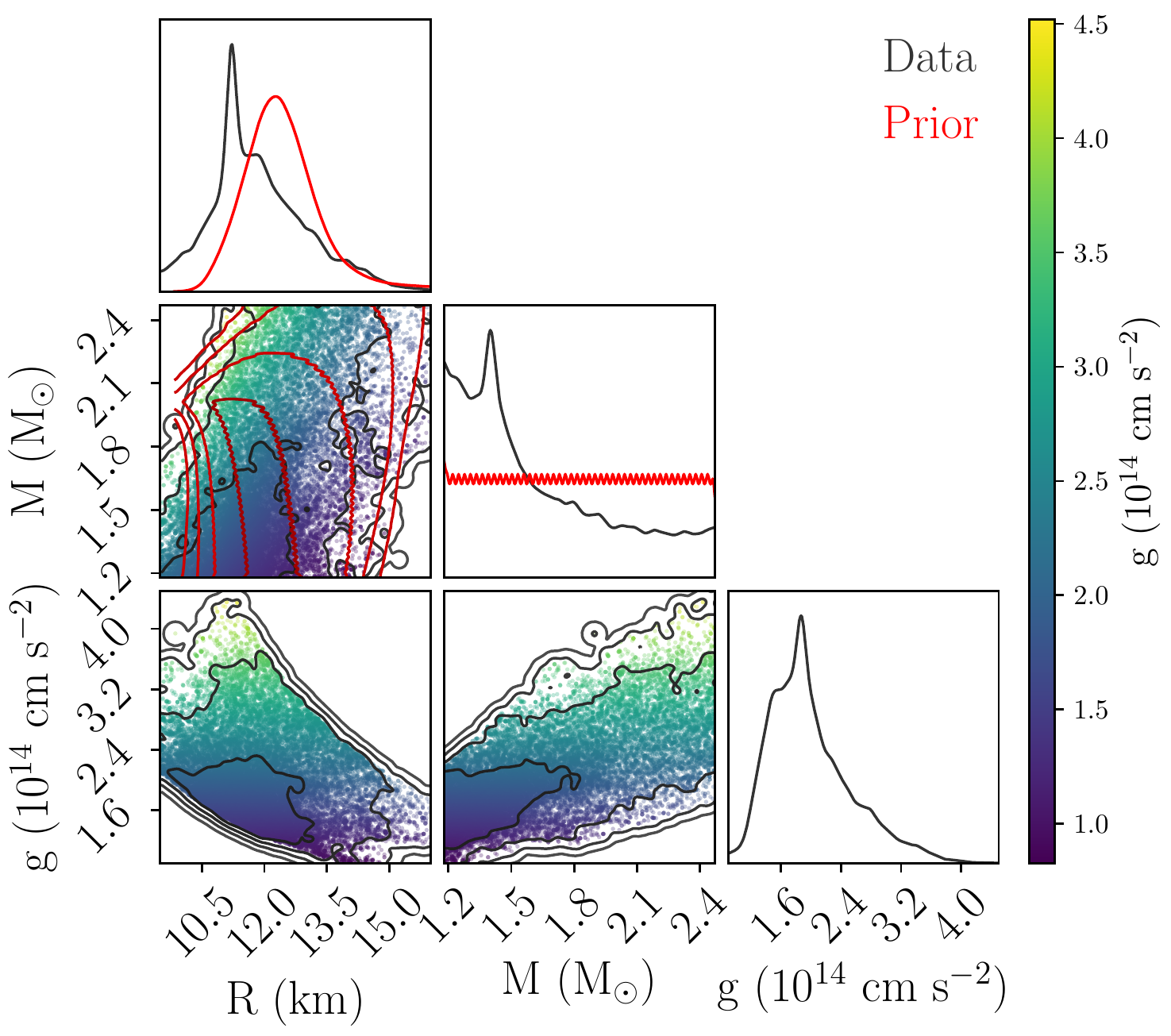}
	\caption{2-dimensional projection of the posterior probabilities of NS mass ($M$), NS radius ($R$) and surface gravity ($g$) of \texttt{SAX J1808.4--3658} (black and coloured by $g$) and the NS mass and radius prior probability distribution from \citet{Steiner2018} (red). Histograms along the diagonal show the marginalised probabilities of the individual parameters and contour levels are $1\,\sigma$ intervals up to $5\,\sigma$.}
    \label{fig:mrresults}
\end{figure}

The posterior distributions and probability contours for mass and radius are plotted in Figure \ref{fig:mrresults}, coloured by surface gravity. The prior we used for mass and radius is plotted in red.

From Figure \ref{fig:mrresults} it appears that there is a preferred surface gravity loosely driving the correlation between mass and radius, with the preferred surface gravity being $1.9\times10^{14}\,\mathrm{g}\,\mathrm{cm}^{-2}$. Interestingly, the radius posterior peaks at a radius lower than the peak radius of the prior distribution, indicating that the model fit prefers a smaller radius. The model is thus sensitive to changes in radius. We explore this further in Section \ref{sec:appendix}, in which we compare model runs with fixed mass and radius and flat mass and radius priors.
\\
\subsection{Binary System Parameters and Inclination}

The mass function of the binary orbit is given by

\begin{equation}
\label{eq:massfunction}
    f_x = \frac{(M_{\mathrm{c}} \mathrm{sin}\,i)^3}{(M_{\mathrm{NS}} + M_{\mathrm{c}})^2} = \frac{4\pi^2(a_x\mathrm{sin}\,i)^3}{GP_{\mathrm{orb}}^2}
\end{equation}

where $i$ is the binary inclination, $M_{\mathrm{c}}$ is the companion star mass and $M_{\mathrm{NS}}$ is the neutron star mass \citep[e.g.,][]{Chakrabarty1998, bildsten2001}. 

Setting the projected semi-major axis, $a_x\,\sin\,i=62.809\,\mathrm{light-ms}$ \citep{Chakrabarty1998} and orbital period, $P_{\mathrm{orb}}=2.01\,\mathrm{h}$ gives $f_x=3.8\times10^{-5}\,\mathrm{M}_{\odot}$. This leaves $M_{\mathrm{NS}}$, $i$, and $M_\mathrm{c}$. Our MCMC algorithm uses {\sc settle} to find $M_{\mathrm{NS}}=1.5^{+0.6}_{-0.3}\,\mathrm{M}_{\odot}$. Given estimates of $M_\mathrm{c}= 0.02-0.011\,\mathrm{M}_{\odot}$ \citep[e.g.,][]{bildsten2001} and using Equation \ref{eq:massfunction}, the mass function gives cos $i=0.62^{+0.09}_{-0.08}$ ($i=\left(64^{+37}_{-20}\right)^\circ$).

Lack of deep X-ray eclipse rules out $i>82^\circ$ ($\cos i < 0.15$) \citep{Chakrabarty1998}.  There is a $2\%$ modulation in X-ray intensity at the orbital period, which suggests that $i$ must be large enough to allow partial X-ray blockage from some circumbinary material \citep{bildsten2001}. Those authors rule out face on inclination due to the 2 hour single-peaked modulation in the optical intensity, confirmed by \citet{homer2001}. The flux minimum occurs when the neutron star is behind the companion, implying that the X-ray heating of the companion is the origin of the modulation, which would have no effect in a face on system. Our best estimate gives $\cos i=0.36^{+0.07}_{-0.04}$ ($i=\left(69^{+4}_{-2}\right)^\circ$), which agrees within $3\,\sigma$ of $\cos i=0.62^{+0.09}_{-0.08}$ found using the mass function, and is within the ranges predicted by \citet{bildsten2001} and \citet{wang2001}. More recently, \citet{Wang2013} and \citet{cackett2009} have attempted to constrain the inclination through optical modelling and iron line modelling respectively. \citet{Wang2013} found $i=\left(50^{+6}_{-5}\right)^\circ$ and \citet{cackett2009} found $i=\left(55^{+8}_{-4}\right)^\circ$ ($1\,\sigma$ uncertainties). The \citet{cackett2009} estimate agrees within $2\,\sigma$ of our best estimate, and the \citet{Wang2013} estimate agrees within $3\,\sigma$.

\begin{figure}
 	\includegraphics[width=\columnwidth,viewport=11 1 455 287,clip]{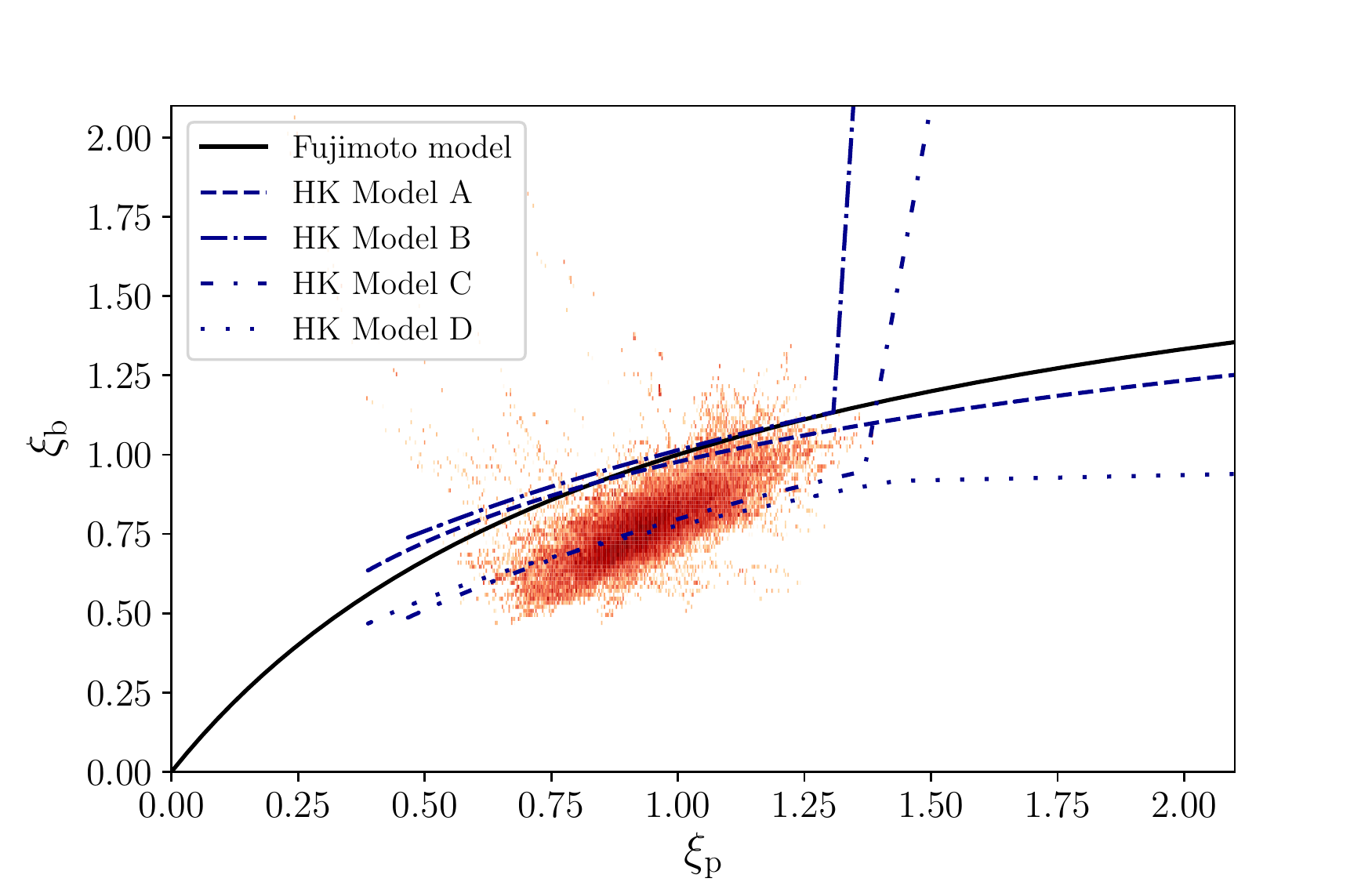}
    \caption{The distribution of the $1\,\sigma$ MCMC predictions for the anisotropy factors, $\xi_{\mathrm{p}}$ and $\xi_{\mathrm{b}}$ (orange) and the relationship between $\xi_{\mathrm{p}}$ and $\xi_{\mathrm{b}}$ as predicted by the \citet{he2016} (HK) and \citet{fujimoto1988} models for different disc shapes.}
    \label{fig:anisotropies}
\end{figure}

Since the anisotropy factors are determined independently in this method, we checked the self consistency of these factors with the relationships predicted by the \citet{fujimoto1988} and \citet{he2016} models in Figure \ref{fig:anisotropies}. 

\texttt{SAX J1808.4--3658} is in the hard spectral state during its outbursts and due to its nature as an accreting pulsar, it is predicted to have a truncated disc \citep[e.g.,][]{Psaltis1999}. The \citet{fujimoto1981} and \citet{he2016} models do not model for a truncated disc, and thus we do not expect any of them to match our predictions perfectly. We could not decisively conclude which of the models matched our predictions best, as they all pass through the $1\,\sigma$ confidence area in Figure \ref{fig:anisotropies}, motivating us to use the simplest model, \citet{he2016} Model~A.

\subsubsection{Wind loss}

Recent research has found that a significant amount ($\approx 30\%$) of the nuclear energy of a Type I X-ray burst could be used to unbind matter from the neutron star surface, ejected in a wind. \citet{Yupaper} found that wind losses can cause up to 30$\%$ loss of nuclear energy in ejecting material from the system. We do not correct for wind loss in our models as upon testing it simply reduced the predicted distance by $\approx 7\%$, with mass loss causing the source to be predicted closer due to fainter bursts being predicted.

\subsection{Chain Convergence and MCMC performance}
\label{sec:chainconvergence}

The convergence of MCMC chains is notoriously difficult to assess. \citet{goodman2010} recommend using the integrated autocorrelation time ($\tau$) to quantify the effects of sampling error on the results. The autocorrelation time provides a measure of the large time error of the Monte Carlo estimator, and so is a reliable indicator of the accuracy of the sampler by measuring the asymptotic variance in the limit of long chains. This essentially measures the effective number of independent samples, and enables estimation of the number of samples required to reduce the relative error on the target integral to a few percent. \citet{Foreman2013} recommend running the chains for at least $50\,\tau$ samples (or conservatively $100\,\tau$ samples).

\begin{figure}
	\includegraphics[width=\columnwidth,viewport=0 0 325 255,clip]{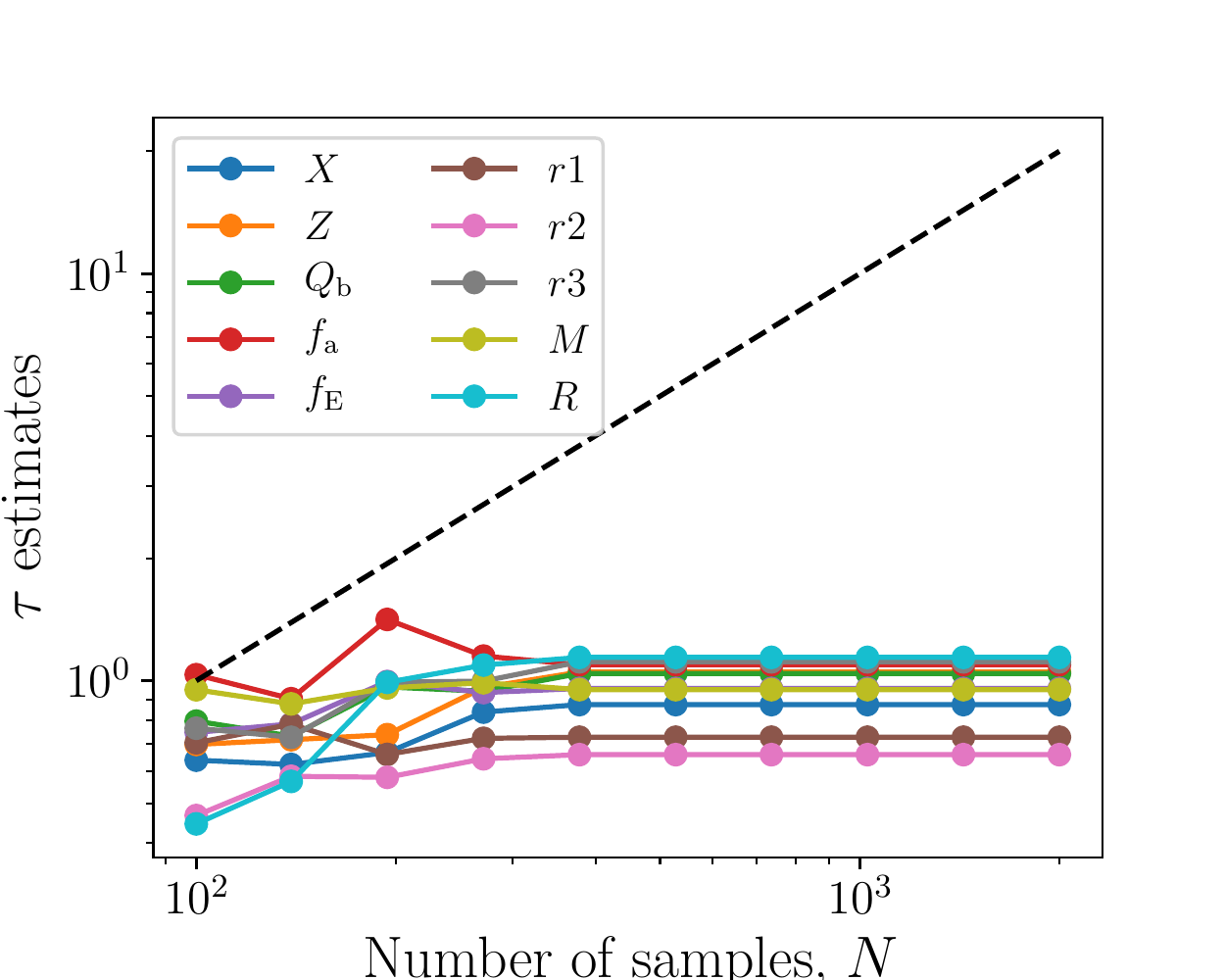}
    \caption{The evolution of the autocorrelation time ($\tau$) with the number of samples taken for the 10 parameters varied with MCMC. The black dashed line is the $\tau$ = $N/100$ line.}
    \label{fig:autocorr}
\end{figure}

In Figure \ref{fig:autocorr} the autocorrelation time of each parameter we varied is plotted as a function of the number of samples. The autocorrelation time is not reliably estimated early on, as there are not enough samples. As more samples are taken, the autocorrelation time estimate becomes more accurate, and after $\approx500$ samples, the autocorrelation time converged on a single value for all of the parameters. By 2,000 samples, all of the autocorrelation estimates are well below the $\tau = N/100$ line, indicating this is an accurate estimate of the autocorrelation time. The largest autocorrelation time is $<$10 so the chains should be converged according to this convergence test after 1,000 steps. We ran the chains for 2,000 steps and 300 walkers to ensure they would be fully converged. 

The acceptance fraction of the samples is a widely used measure to check the performance of MCMC. The acceptance fraction is the number of proposed steps that are accepted, and will be small if too many steps are rejected and large if too many steps are accepted \citep{Foreman2013}. The general consensus is that the acceptance fraction should be between $0.2$--$0.5$ \citep[e.g.,][]{Gelman1996}, however, it depends on the model and situation. The acceptance fraction of our final run was $0.14$, which is lower than $0.2$ but not unreasonably low. As a test, we ran the MCMC code returning only the prior (excluding the likelihood and thus model call) and found an acceptance fraction of $0.11$.  We can thus conclude that the prior limits are constraining parameter space and causing a larger than usual amount of steps to be proposed outside the allowed domain and be rejected. 

A final test we ran to ensure the end walker positions were independent of the initial positions was to change the size of the Gaussian ball the walkers were initialised around by a factor of 100. This did not affect the marginalised posterior distributions of each parameter. 

\section{Discussion and Conclusions}\label{sec:conclusion}

We have used a Bayesian approach to successfully match observed burst fluence, alpha, and recurrence times with the {\sc settle} model to infer system parameters of an accreting millisecond pulsar in outburst. The properties of the bursts observed from \texttt{SAX J1808.4--3658} indicate they arrive in helium rich fuel, given the relatively short length, low flux, and high alpha values \citep[e.g.,][]{galloway2006}. We inferred $\bar{X} < 0.2$ for all bursts, as expected for helium rich bursts. Overall, we matched the observed burst fluences, alphas, and recurrence times to within the $1\,\sigma$ uncertainty of the model. 

Based on the agreement between the observed burst properties, we then inferred and modelled composition, base flux, neutron star mass, neutron star radius, distance, and inclination. We set the prior for metallicity to the expected distribution of metallicities based on the location of \texttt{SAX J1808.3--3658} in the Galaxy, and deduced the most likely CNO mass fraction (metallicity) of the source to be between 0.009--0.019. This range covers the expected value for solar CNO metallicity, which recent work puts at $0.01$ \citep{Lodders2009}. We inferred a slightly depleted hydrogen mass fraction of the accreted fuel, of $0.58^{+0.13}_{-0.14}$, indicating that the companion star in this system could be significantly evolved. 

We inferred a base flux of $0.25$--$0.7\,\mathrm{MeV}/\mathrm{nucleon}$, which gives an indication of the base heating of the neutron star. We used the constraints on the relationship between NS mass and radius calculated by \citet{Steiner2018} for neutron stars in low mass X-ray binaries in globular clusters as an informed prior for the distribution of acceptable mass and radius combinations and inferred a NS mass between $0.9$--$1.8\,\mathrm{M}_{\odot}$ and a radius between $10.8$--$13.1\,\mathrm{km}$.  The mass is constrained to within $40\%$ ($1\,\sigma$ limit) and the radius is constrained to within $11\%$ ($1\,\sigma$ limit), indicating that this method cannot constrain neutron star mass and radius to a high precision. This is most likely due to degeneracies in some of the other model parameters, such as base heating ($Q_{\mathrm{b}}$), that can adjust to produce the same burst energies and recurrence time when the mass, radius, and thus surface gravity and redshift are varied. 

Finally, we used the scaling factors we defined and the posterior distributions of the parameters to infer a distance and inclination of \texttt{SAX J1808.4--3658}, based on \citet{he2016} disc Model~A. We found a distance range of $3.1$--$3.6\,\mathrm{kpc}$ for an inclination from the rotation axis of $67^\circ$--$73^\circ$.  \citet{galloway2006} used the fact that the observed bursts showed photospheric radius expansion to infer an upper limit of the distance to the source of $3.6\,\mathrm{kpc}$, just within our $1\,\sigma$ upper limit on the distance. An inclination of $69^\circ$, implies both the observed burst and persistent fluxes are artificially increased by reflection from the disc and by the disc itself, bringing the source closer when compared to distance estimates such as \citet{galloway2006} that do not account for inclination. 

We have reported the median value and 68th percentile limits of the predicted posterior distributions to give an indication of the parameter values. While this is a representation of the data, the true solutions are the posterior distributions that were calculated by the MCMC algorithm. Parameters estimated using this technique are limited to the accuracy of the {\sc settle} model predictions. The benefit of using such a simple model is that it can be used in a MCMC algorithm to effectively sample parameter space.

We have demonstrated that we can infer system parameters of a bursting source using observations of an outburst and the {\sc settle} model. Future work could apply this method of matching observed properties with models to the other known bursting AMSPs with observations of burst trains, or any accreting neutron star with observations of burst trains. This method enables initial exploration of parameter space, to determine an estimate for the range of parameters for each source. These ranges could then be used in more advanced, computationally expensive models such as \kepler to narrow down the parameter limits.

\section*{Acknowledgements}

We thank Andrew Casey for his contribution of the metallicity prior we used. We thank Andrew Steiner for helpful discussions regarding the neutron star mass-radius prior. This paper utilizes preliminary analysis results from the Multi-INstrument Burst ARchive (MINBAR), which is supported under the Australian Academy of Science's Scientific Visits to Europe program, and the Australian Research Council's Discovery Projects and Future Fellowship funding schemes. AG and ZJ acknowledge support by an Australian Government Research Training (RTP) Scholarship.  AH was supported by a grant from Science and Technology Commission of Shanghai Municipality (Grant No.~16DZ2260200) and National Natural Science Foundation of China (Grant No.~11655002). This work benefited from support by the National Science Foundation under Grant No. PHY-1430152 (JINA Center for the Evolution of the Elements). We thank the anonymous referee for constructive and insightful comments that helped to improve the manuscript.

\bibliographystyle{mnras}
\bibliography{bibfile} 

\clearpage
\appendix
\label{sec:appendix}

\section{Additional Model Runs}

In this section we provide the results of additional model runs carried out in order to robustly test the MCMC code, as well as to replicate past works. 

\subsection{Fixed mass and radius}

\begin{figure*}
 	\includegraphics[width=2\columnwidth]{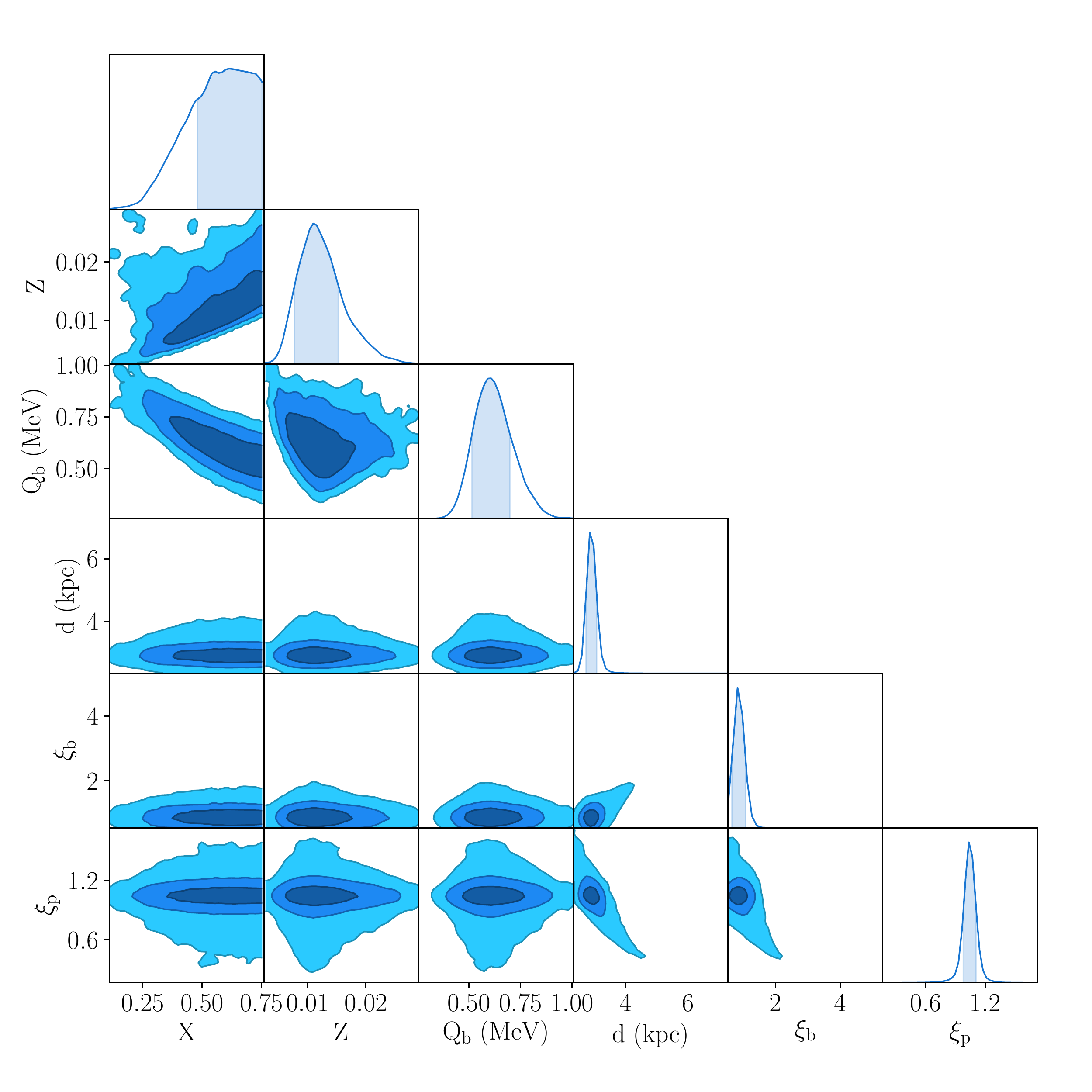}
    \caption{Marginalised probability distributions for hydrogen fraction $X$, CNO metallicity $Z$, base flux $Q_{\mathrm{b}}$, distance $d$ and anisotropy factors $\xi_\mathrm{p}$ and $\xi_\mathrm{b}$ for the MCMC run with fixed NS mass of $1.4\,\mathrm{M}_{\odot}$ and radius of $11.2\,\mathrm{km}$.
    \label{fig:fixedmrwalks}}
\end{figure*}

We held mass and radius constant at $M = 1.4\,$M$_{\odot}$ and $R = 11.2\,$km to replicate the analysis of \citet{galloway2006}. The parameter predictions are presented in Table \ref{tab:fixedmr_parameters} and the probability contours for X, Z, Q$_\mathrm{b}$, d, $\xi_{\mathrm{p}}$ and $\xi_{\mathrm{b}}$ are presented in Figure \ref{fig:fixedmrwalks}. 

{
\renewcommand{\arraystretch}{1.6}
\begin{table}
	\centering
	\caption{\texttt{SAX J1808.4--3658} derived neutron star parameters for a fixed mass and radius}
	\label{tab:fixedmr_parameters}
	\begin{tabular}{lccr}
		\hline
		Parameter & Value\\
		\hline
		$X$ & $0.57^{+0.13}_{-0.15}$ \\
		$Z$ & $0.012^{+0.004}_{-0.003}$ \\
		$Q_{\mathrm{b}}$ ($\mathrm{MeV}/\mathrm{nucleon}$) & $0.6^{+0.1}_{-0.1}$ \\
		$M$ ($\mathrm{M}_{\odot}$) & $1.4$ \\
		$R$ ($\mathrm{km}$) & $11.2$\\
		$\dot m_\mathrm{max}$ ($\dot{m}_\mathrm{Edd}$)&$0.035^{+0.002}_{-0.001}$ \\
		$g$ ($10^{14}\,\mathrm{cm}\,\mathrm{s}^{-2}$)& $1.86$ \\
		$1+z$ & $1.26$ \\
		$d$ ($\mathrm{kpc}$) & $2.9^{+0.1}_{-0.1}$ \\
		$\xi_{\mathrm{b}}$ & $0.9^{+0.1}_{-0.1}$ \\
		$\xi_{\mathrm{p}}$ & $1.04^{+0.05}_{-0.04}$ \\
		$\cos i$ & $0.35^{+0.08}_{-0.05}$\\
		\hline
	\end{tabular}
	
\end{table}
	}
	
The predicted hydrogen mass fraction in Table \ref{tab:fixedmr_parameters} matches within uncertainty of those predicted by \citet{galloway2006} ($X_{\mathrm{0}} = 0.54$ for $Z = 0.02$, or $X_{\mathrm{0}} = 0.5$ for $Z = 0.016$). \citet{galloway2006} inferred a base flux of $Q_{\mathrm{b}} = 0.325$ and a distance of 3.1--3.8 kpc which is just outside the lower limit of our 1-$\sigma$ range of predictions for these parameters. \citet{galloway2006} used a chi-squared minimisation approach to determine the parameters of interest given a fixed grid of parameters in $X$, $Z$ and $Q_{\mathrm{b}}$ space to find the best fit burst fluence, recurrence times and alpha values. We used a more robust approach to determine the parameters of interest, and vary more parameters than just $X$, $Z$, and $Q_{\mathrm{b}}$ and so we do not expect our results to match exactly with those of \citet{galloway2006}. The composition constraints and distance estimate broadly agree with those found by \citet{galloway2006} for a fixed NS mass and radius. 

\subsection{Flat priors on mass and radius}

\begin{figure*}
 	\includegraphics[width=2\columnwidth]{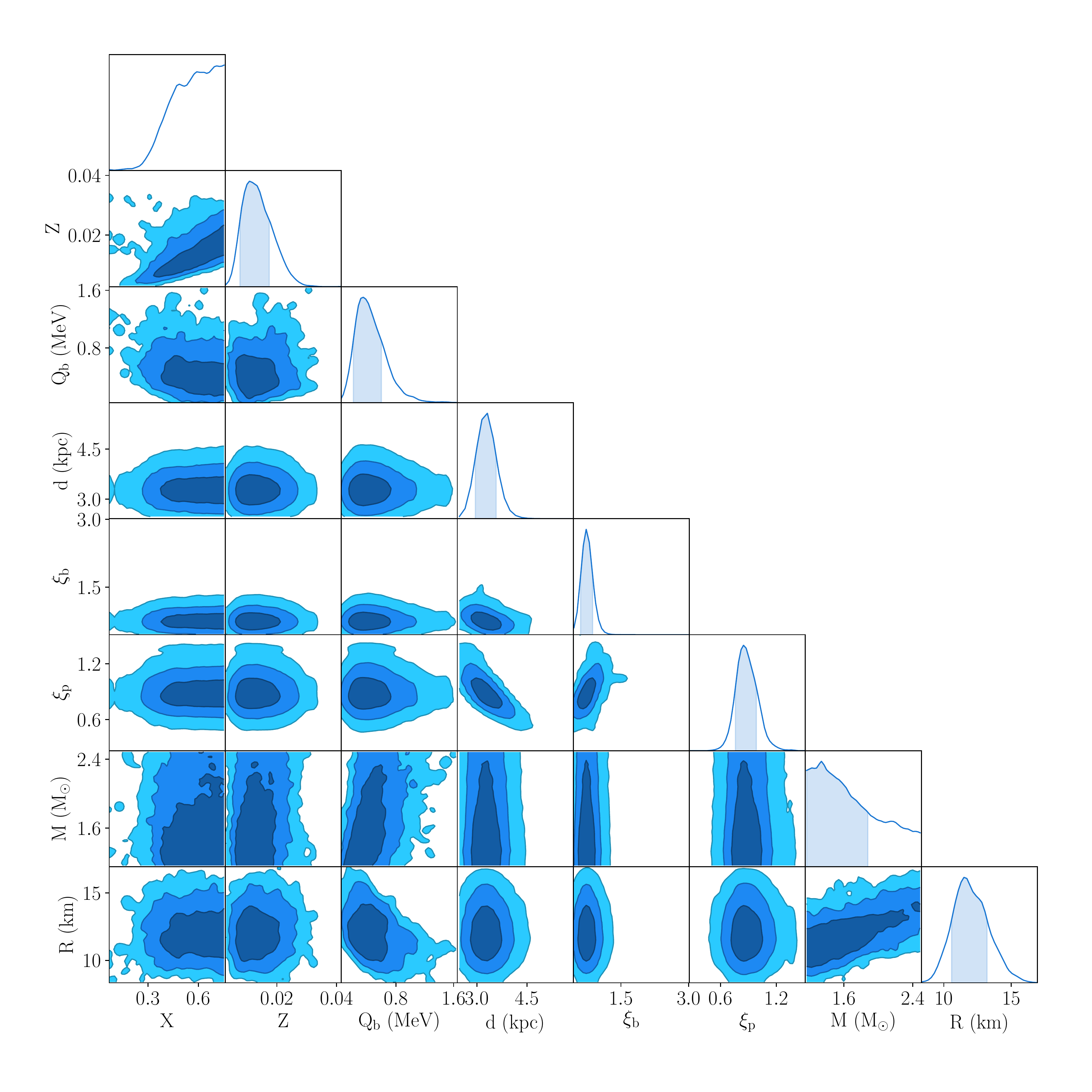}
    \caption{Marginalised probability distributions for hydrogen fraction $X$, CNO metallicity $Z$, base flux $Q_{\mathrm{b}}$, distance $d$ and anisotropy factors $\xi_\mathrm{p}$ and $\xi_\mathrm{b}$, NS mass $M$, and NS radius $R$ for the MCMC run with flat prior ranges for $M$ and $R$.}
    \label{fig:flatmrpriorwalks}
\end{figure*}

Here we show the parameter limits when assuming a flat prior for mass and radius, rather than the \citet{Steiner2018} probability distribution. The parameter limits are reported in Table \ref{tab:flatmrpriorparameters} and posterior distributions are plotted in Figure \ref{fig:flatmrpriorwalks}. 
{
\renewcommand{\arraystretch}{1.6}
\begin{table}
	\centering
	\caption{\texttt{SAX J1808.4--3658} derived neutron star parameters}
	\label{tab:flatmrpriorparameters}
	\begin{tabular}{lccr}
		\hline
		Parameter & Value\\
		\hline
		$X$ & $0.58^{+0.13}_{-0.15}$ \\
		$Z$ & $0.013^{+0.006}_{-0.004}$ \\
		$Q_{\mathrm{b}}$ ($\mathrm{MeV}/\mathrm{nucleon}$) & $0.4^{+0.2}_{-0.2}$ \\
		$M$ ($\mathrm{M}_{\odot}$) & $1.6^{+0.5}_{-0.3}$ \\
		$R$ ($\mathrm{km}$) & $11.9^{+1.4}_{-1.2}$ \\
		$\dot{m}_{\mathrm{max}}$  ($\dot{m}_\mathrm{Edd}$)&$0.037^{+0.002}_{-0.002}$ \\
		$g$ ($10^{14}\,\mathrm{cm}\,\mathrm{s}^{-2}$)& $1.95^{+0.6}_{-0.4}$ \\
		$1+z$ & $1.29^{+0.1}_{-0.06}$ \\
		$d$ ($\mathrm{kpc}$) & $3.3^{+0.2}_{-0.2}$ \\
		$\xi_{\mathrm{b}}$ & $0.74^{+0.1}_{-0.10}$ \\
		$\xi_{\mathrm{p}}$ & $0.87^{+0.1}_{-0.10}$ \\
		$\cos i$ & $0.37^{+0.07}_{-0.05}$\\
		\hline
	\end{tabular}
		
\end{table}
}

When assuming a flat prior range in mass and radius compared to assuming a more informative prior, we find that the mass is equally well constrained in both cases but that the radius is better constrained for the more informative prior run. Otherwise, all parameters remain approximately the same, except for surface gravity, with a stronger surface gravity required by the flat prior run.

\bsp
\label{lastpage}
\end{document}